\documentclass[letterpaper]{article} 
\usepackage{aaai2027}  
\usepackage[hyphens]{url}  
\usepackage{graphicx} 
\usepackage{natbib}  
\AtBeginDocument{%
  \let\OriginalAAAICite\cite
  \renewcommand{\cite}[1]{%
    \if\relax\detokenize{#1}\relax
    \else
      \OriginalAAAICite{#1}%
    \fi}}
\usepackage{caption} 
\usepackage{amsmath}
\usepackage{amssymb}
\usepackage{algorithm}
\usepackage{algpseudocode}
\usepackage{array}
\usepackage{placeins}

\usepackage{newfloat}
\usepackage{listings}
\DeclareCaptionStyle{ruled}{labelfont=normalfont,labelsep=colon,strut=off} 
\floatstyle{ruled}
\newfloat{listing}{tb}{lst}{}
\floatname{listing}{Listing}

\usepackage{booktabs}

\newcommand{\method}{AnyGS2Mesh}

\title{AnyGS2Mesh: Feed-Forward Mesh Reconstruction from 3D Gaussian Splatting with Arbitrary-Resolution Views}
\author{
Yuxuan Song,
Fan Gao,
Yibo Zhao,
Jiarui Wen,
Youcheng Cai,
Ligang Liu
}

\affiliations{
University of Science and Technology of China, Hefei, China
}

\begin{document}

\maketitle

\begin{abstract}
Existing 3D mesh reconstruction methods from Gaussian scene representations predominantly rely on iterative optimization, resulting in slow inference and limited scalability to high-resolution inputs. In this paper, we present AnyGS2Mesh, the first feed-forward framework for directly reconstructing 3D meshes from 3D Gaussian Splatting representations with support for arbitrary input image resolutions. Our approach incorporates a Gaussian-Guided Transformer architecture that exploits explicit 3D geometric priors for efficient mesh generation. We introduce three key components: (1) a Gaussian-Guided Spatial Reasoning Transformer represents Gaussian primitives as structured 3D tokens and jointly reasons over Gaussian and image features; (2) a Streaming and Patchwise Geometry Encoder processes native-resolution views sequentially and aggregates information across variable-length view sets; (3) a Scale-Aligned Hybrid Depth Refiner uses a PatchFusion-style encoder--decoder to fuse RGB-conditioned predicted depth with Gaussian-rendered metric depth, combining fine local structures with globally consistent metric scale. The refined depth maps are integrated through TSDF fusion, followed by Marching Cubes for deterministic mesh extraction. Extensive experiments show that AnyGS2Mesh achieves state-of-the-art reconstruction quality while significantly reducing inference time compared with optimization-based baselines, enabling near-real-time,  high-quality mesh reconstruction. Our results demonstrate the potential of combining Gaussian representations and feed-forward Transformer architectures for scalable 3D geometry reconstruction. The code will be made publicly available upon acceptance.
\end{abstract}


\section{Introduction}

Reconstructing high-quality 3D meshes from multi-view observations has long been a fundamental problem in computer vision and computer graphics, with broad applications in virtual and augmented reality \cite{}, autonomous driving \cite{}, robotic perception \cite{}, and physical simulation \cite{}. Recently, 3D Gaussian Splatting (3DGS) \cite{kerbl20233dgs} has emerged as a powerful scene representation for novel view synthesis. By optimizing a set of anisotropic Gaussian primitives, 3DGS achieves high-fidelity rendering while maintaining efficient training
\noindent\begin{minipage}{\columnwidth}
    \centering
    \includegraphics[width=\columnwidth]{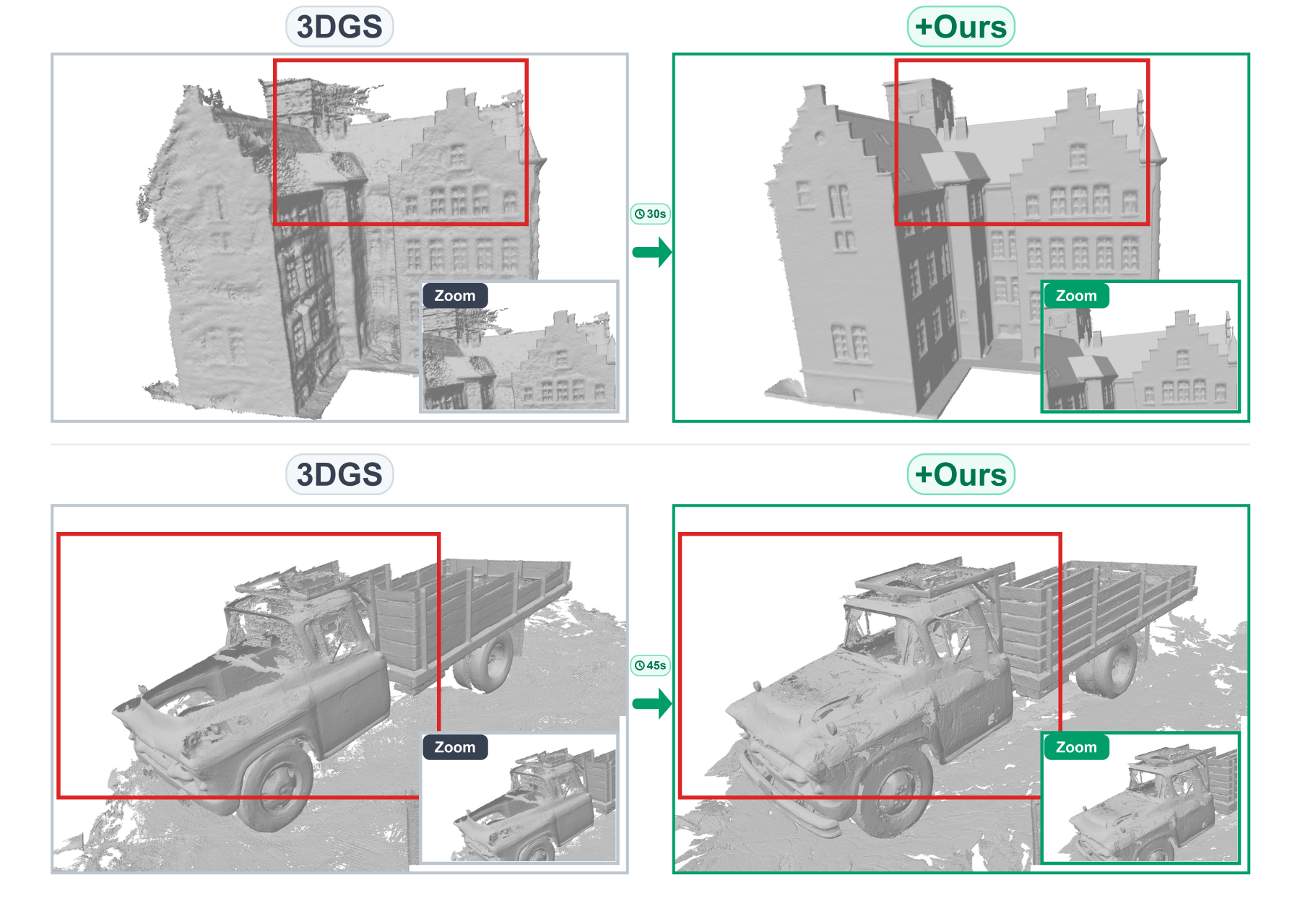}
    \captionof{figure}{Qualitative mesh reconstruction results of AnyGS2Mesh on the DTU and Tanks and Temples datasets.}
    \label{fig:teaser}
\end{minipage}
\par
\vspace{1em}
and real-time rendering performance. Nevertheless, most subsequent studies have primarily focused on improving rendering quality \cite{}, because the optimized Gaussian centers and covariances do not necessarily form a coherent surface. As a result, the direct extraction of high-quality meshes from 3DGS remains challenging.

To improve mesh reconstruction quality, existing methods either incorporate stronger geometric priors into Gaussian primitives or introduce additional geometric supervision. For instance, SuGaR \cite{guedon2024sugar} encourages Gaussian primitives to align with local surface geometry, 2DGS \cite{huang2024twodgs} constrains volumetric Gaussian primitives to oriented planar disks, and QGS \cite{zhang2025qgs} further replaces first-order planar primitives with second-order quadric surfaces to better model high-curvature regions. These methods are capable of producing high-quality meshes, yet they continue to rely on scene-specific optimization.

As 3DGS assets are becoming increasingly prevalent for novel view synthesis, directly converting these rendering-oriented Gaussian representations into high-quality meshes has emerged as an important research problem. GS2Mesh \cite{wolf2024gs2mesh} takes an initial step in this direction by rendering stereo-aligned novel views from 3DGS and estimating depth using a pre-trained stereo-matching model. However, it still relies on a heuristic stereo depth fusion pipeline and does not fully exploit the explicit geometric information encoded in Gaussian primitives.

To overcome the inefficiency of per-scene optimization and the limitations of heuristic fusion strategies, feed-forward reconstruction provides a promising alternative. Recent transformer-based architectures have demonstrated strong capabilities in geometric reasoning \cite{tang2023vggt,wang2024mvsformer}. In particular, VGGT \cite{tang2023vggt} demonstrates that transformers can infer geometric representations from arbitrary input views in an end-to-end manner, thereby improving inference efficiency and generalization performance. Our key idea is to extend feed-forward reasoning to Gaussian-to-mesh reconstruction by treating Gaussian primitives as structured 3D tokens. However, two challenges remain: (1) how to effectively integrate the explicit geometric priors encoded in Gaussian primitives into transformer-based architectures, and (2) how to support multi-view inputs with arbitrary numbers of images and varying image resolutions. In practical applications, the synthesized novel views from 3DGS often vary in both number and resolution, whereas conventional vision transformers typically require fixed-length token sequences.

To address these challenges, we propose AnyGS2Mesh, a novel feed-forward framework for the direct reconstruction of high-quality 3D meshes from 3DGS representations, supporting arbitrary numbers of input images at arbitrary resolutions. First, we introduce a \textbf{Gaussian-Guided Spatial Reasoning Transformer (GSRT)} that represents Gaussian primitives as structured 3D tokens and jointly reasons over Gaussian and image tokens for multi-view geometric reasoning. Second, we propose a \textbf{Streaming and Patchwise Geometry Encoder (SPGE)} that processes input views in a streaming manner through patchwise tokenization and feature aggregation, enabling inference from arbitrary numbers and arbitrary resolution of input images without requiring fixed-size resizing or cropping. Third, we design a \textbf{Scale-Aligned Hybrid Depth Refiner (SHDR)} that combines detail-preserving relative depth prediction with globally scale-consistent Gaussian-derived depth. Extensive experiments across multiple benchmark datasets, including DTU \cite{jensen2014dtu}, Mip-NeRF 360~\cite{barron2022mipnerf360}, and Tanks and Temples \cite{knapitsch2017tanks}, demonstrate that AnyGS2Mesh achieves reconstruction quality comparable to state-of-the-art optimization-based approaches, while reducing inference time by a significant margin.

Our main contributions are summarized as follows:
\begin{itemize}
\item We propose \textbf{AnyGS2Mesh}, a feed-forward Gaussian-to-mesh reconstruction framework that directly converts 3D Gaussian Splatting representations into explicit meshes, avoiding costly per-scene optimization.
\item We introduce a \textbf{GSRT} that treats Gaussian primitives as structured 3D tokens and jointly reasons over Gaussian and image features, enabling explicit 3D priors to guide multi-view geometric reconstruction.
\item We design a \textbf{SPGE} together with a \textbf{SHDR}, enabling arbitrary numbers and arbitrary resolution of input images while improving both local geometric details and global depth consistency.
\end{itemize}

\section{Related Work}

\subsection{Gaussian Splatting-based Surface Reconstruction}

3DGS represents a scene with anisotropic primitives carrying geometry, opacity, and appearance, and renders them efficiently through differentiable splatting \cite{kerbl20233dgs}. Its explicit spatial structure and real-time rendering have made it a common representation for scene modeling, mapping, and novel-view synthesis \cite{chen2024gs_survey,keetha2023splatam,yugay2023gaussian_slam}. However, photometric fitting alone does not force the primitives to form a surface.

Surface-oriented methods address this mismatch by modifying the representation or its optimization. SuGaR aligns Gaussians with local surfaces \cite{guedon2024sugar}; 2DGS and QGS use more surface-oriented primitives \cite{huang2024twodgs,zhang2025qgs}; RaDe-GS and PGSR improve depth or planar consistency \cite{zhang2024radegs,chen2024pgsr}; and GausSurf adds explicit geometry guidance \cite{wang2024gaussurf}. These approaches obtain strong geometry but remain coupled to scene-specific fitting. GS2Mesh instead converts an existing 3DGS through rendered stereo views \cite{wolf2024gs2mesh}. Our work follows this post-hoc setting while exposing primitive-level Gaussian attributes directly to a learned geometric reasoner.

\subsection{Feed-forward 3D Reconstruction}

Transformer-based models increasingly amortize multi-view reasoning across scenes. MVSFormer learns global multi-view interactions for depth estimation, while VGGT jointly predicts cameras, depth maps, and point maps in a feed-forward architecture \cite{wang2024mvsformer,tang2023vggt}. OmniVGGT extends this paradigm with depth and camera priors, and AnySplat connects transformer-based geometry prediction with Gaussian scene generation \cite{peng2025omnivggt,jiang2025anysplat}.

Resolution-flexible encoders such as NaViT and Any Resolution Any Geometry process variable image shapes through patch-based representations \cite{dehghani2023navit,cui2026anyresanygeo}. PatchFusion targets high-resolution depth, while STream3R processes multi-view geometry sequentially \cite{li2024patchfusion,lan2025stream3r}. These methods address image scale or sequence length, but do not combine native-resolution streaming with primitive-level 3DGS guidance for post-hoc mesh conversion.

Existing feed-forward methods mainly predict intermediate geometry from images or construct a new scene representation. They do not directly solve post-hoc conversion of an existing 3DGS into a mesh, nor do they jointly address primitive-level 3D guidance, native-resolution visual encoding, and metric depth alignment. AnyGS2Mesh targets this intersection with three corresponding modules.

\section{Preliminaries}
\paragraph{3DGS.}
3D Gaussian Splatting (3DGS) \cite{kerbl20233dgs} models a scene using a set of anisotropic Gaussian primitives $\mathcal{G}=\{\mathbf{g}_k\}_{k=1}^{K}$. Each primitive is parameterized by a center $\mathbf{p}_k$, a 3D covariance matrix $\boldsymbol{\Sigma}_k$, an opacity $\alpha_k$, and a color $\mathbf{c}_k$, and is defined as, 

\begin{equation}
G_k(\mathbf{p}) = \exp \left(
-\frac{1}{2}(\mathbf{p}-\mathbf{p}_k)^T \boldsymbol{\Sigma}_k^{-1} (\mathbf{p}-\mathbf{p}_k)
\right),
\end{equation}
where $\boldsymbol{\Sigma}_k$ is typically parameterized by learnable scale and rotation parameters.

Given a camera view, each primitive is projected onto a 2D Gaussian footprint $G_k^{2D}$. The projected primitives are sorted according to their camera-space depth and composited via alpha blending to obtain the color of pixel $\mathbf{u}$:

\begin{equation}
I(\mathbf{u})=\sum_{k=1}^{K}
\mathbf{c}_k \alpha_k G_k^{2D}(\mathbf{u})
\prod_{j=1}^{k-1}\left(1-\alpha_j G_j^{2D}(\mathbf{u})\right).
\end{equation}

\paragraph{VGGT.}
The Visual Geometry Grounded Transformer (VGGT)~\cite{tang2023vggt} demonstrates that transformer architectures can amortize geometric reasoning across unconstrained multi-view inputs. Given $T$ input frames, an image encoder first extracts visual tokens $\mathbf{F}_t$ for each frame. A global multi-view transformer then exchanges information across all frames to produce geometry-aware tokens:
\begin{equation}
\{\mathbf{H}_t\}_{t=1}^{T}
=\Phi_{\mathrm{mv}}\left(\{\mathbf{F}_t\}_{t=1}^{T}\right).
\end{equation}

Task-specific heads decode these tokens into per-frame camera parameters, depth maps, and point maps:
\begin{equation}
\hat{C}_t, \hat{D}_t, \hat{P}_t
=\Psi(\mathbf{H}_t).
\end{equation}

This feed-forward formulation avoids scene-specific optimization and provides a useful template for our Gaussian-to-mesh setting. Different from image-only reconstruction, however, AnyGS2Mesh must additionally exploit the metric 3D structure already encoded in the input Gaussian primitives.

\begin{figure*}[t]
    \centering
    \includegraphics[width=\textwidth]{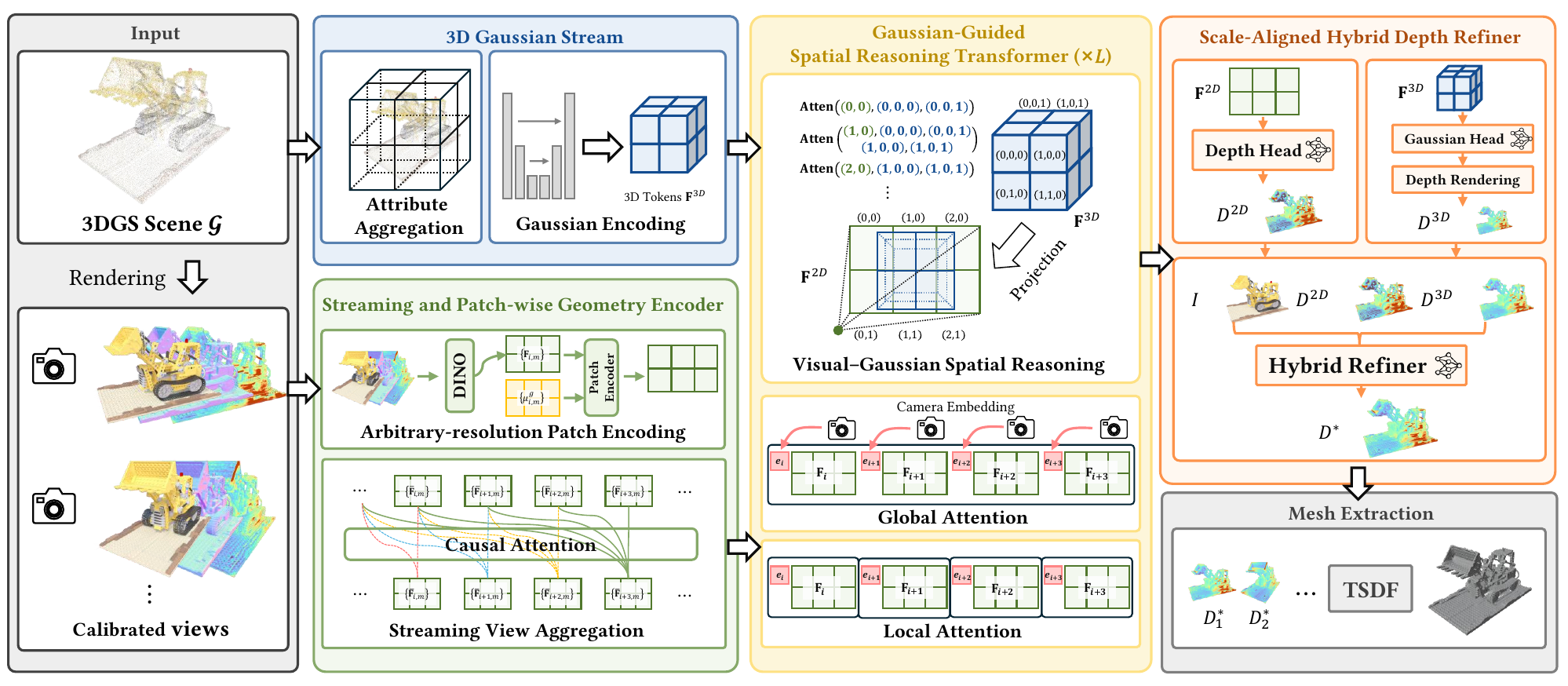}
    \caption{Overview of AnyGS2Mesh. Calibrated views are encoded at native resolution by the Streaming and Patchwise Geometry Encoder. The Gaussian-Guided Spatial Reasoning Transformer aligns those image tokens with PTV3 features from the input 3DGS representation. The Scale-Aligned Hybrid Depth Refiner combines the fused prediction with Gaussian metric depth, after which TSDF fusion and Marching Cubes produce the mesh.}
    \label{fig:framework}
\end{figure*}

\section{Method}
\label{sec:methodology}

AnyGS2Mesh is a feed-forward framework for the direct conversion of an input 3DGS representation into an explicit mesh. As shown in Fig.~\ref{fig:framework}, the pipeline consists of three learned modules. The \textbf{GSRT} transforms irregular Gaussian primitives into view-aligned geometric tokens and leverages them to guide multi-view feature learning. The \textbf{SPGE} preserves native-resolution visual details while aggregating information across variable-length view sequences. The \textbf{SHDR} fuses predicted depth with Gaussian-derived depth to produce scale-aligned depth maps.

\paragraph{Problem Formulation.}
For each scene, given a 3DGS representation
$\mathcal{G}=\{\mathbf{g}_k\}_{k=1}^{K}$ and a set of $N$
calibrated cameras, we render the corresponding RGB views
$\{I_i\}_{i=1}^{N}$ from $\mathcal{G}$ using the standard Gaussian
splatting renderer. Our goal is to learn a scene-agnostic mapping
\begin{equation}
\{D_i\}_{i=1}^{N}
=
f_{\theta}\left(\mathcal{G},\{I_i\}_{i=1}^{N}\right),
\end{equation}
where $D_i$ denotes the predicted depth map for view $i$. The predicted depth maps are then integrated by TSDF fusion \cite{curless1996volumetric} to extract the final mesh. Both the number of input views $N$ and their spatial resolutions may vary across scenes.

\subsection{Gaussian-Guided Spatial Reasoning Transformer}
\label{subsec:gaussian_reasoner}

The Gaussian-Guided Spatial Reasoning Transformer (GSRT) performs explicit information exchange between 2D visual tokens and 3D Gaussian tokens. GSRT therefore maintains two coupled streams: a 2D stream that reasons over multi-view image tokens, and a 3D stream that encodes Gaussian primitives into compact geometry tokens. The two streams interact through view-aligned spatial attention, allowing image features to query the Gaussian structure at their projected spatial support.

\paragraph{2D Visual Stream.}
For each calibrated view, we initialize image tokens with DINOv2~\cite{oquab2023dinov2} and follow the VGGT-style architecture~\cite{tang2023vggt} to encode them with alternating attention blocks. Specifically, image tokens are augmented with camera embeddings, for which we adopt the camera encoding strategy from OmniVGGT~\cite{peng2025omnivggt}. Intra-view attention refines local appearance and geometry cues within each frame, while inter-view attention exchanges information across all available views to enforce multi-view consistency:
\begin{equation}
\mathbf{F}^{2D}
=
\Phi_{\mathrm{2D}}\left(
\{\mathbf{F}_{i},\mathbf{e}_{i}\}_{i=1}^{N}
\right),
\label{eq:gsrt_2d_stream}
\end{equation}
where $\mathbf{F}_{i}$ denotes the image tokens of view $i$, $\mathbf{e}_{i}$ is the corresponding camera embedding, and $\mathbf{F}^{2D}=\{\mathbf{f}^{2D}_{im}\}$ denotes the resulting multi-view visual tokens.

\paragraph{3D Gaussian Stream.}
The input 3DGS representation contains irregularly distributed primitives, making dense transformer reasoning inefficient. We voxelize the Gaussian primitives and aggregate the attributes within each occupied voxel, including position, anisotropic scale, rotation, opacity, and appearance features, into a representative Gaussian token. A Point Transformer V3 (PTV3) backbone~\cite{wu2024pointtransformerv3} then encodes the sparse Gaussian token set into context-aware 3D tokens:
\begin{equation}
\mathbf{F}^{3D}
=
\Phi_{\mathrm{PTV3}}(\mathcal{G}), 
\end{equation}
where $\mathbf{F}^{3D}=\{\mathbf{f}^{3D}_k\}$ denotes the resulting 3D tokens and $\mathbf{f}^{3D}_k$ captures both local and global spatial context.

\paragraph{Visual--Gaussian Spatial Reasoning.}
To exchange information between the two streams, Gaussian tokens are first projected into each calibrated view. For each image token, we collect the visible 3D Gaussian tokens whose projections fall within the corresponding patch region. We then perform local 2D--3D attention, using the image token and its aligned Gaussian tokens as a joint reasoning set:
\begin{equation}
(\hat{\mathbf{f}}^{2D}_{im},\{\hat{\mathbf{f}}^{3D}_{k}\})
=
\operatorname{SelfAttn}
\left(
\mathbf{f}^{2D}_{im},
\{\mathbf{f}^{3D}_{k}\}
\right).
\end{equation}
This operation is inserted at intermediate transformer layers, enabling 2D features to absorb explicit metric geometry while updating the corresponding 3D tokens with view-dependent visual evidence. Applying this reasoning over all views produces fused 2D tokens $\hat{\mathbf{F}}^{2D}$ and updated 3D Gaussian tokens $\hat{\mathbf{F}}^{3D}$ for subsequent decoding.

\paragraph{Feature Decoding.}
We use separate decoding heads for the visual and Gaussian streams. For view $i$, the 2D depth head takes the fused visual tokens $\hat{\mathbf{F}}^{2D}_{i}$ as input and predicts a depth map:

\begin{equation}
D^{2D}_{i}
=
\operatorname{DepthHead}(\hat{\mathbf{F}}^{2D}_{i}),
\label{eq:visual_depth}
\end{equation}
where $D^{2D}_{i}$ denotes the depth prediction from the visual stream, which preserves fine-grained local geometric details but may not be metrically aligned. 

In parallel, the Gaussian head decodes the updated 3D tokens $\hat{\mathbf{F}}^{3D}$ into surface-aware Gaussian primitives, which are then rendered to produce a metric depth map:

\begin{equation}
D_i^{3D}
=
\mathcal{R}_{\mathrm{depth}}
\left(
\operatorname{GaussianHead}(\hat{\mathbf{F}}^{3D})
\right),
\label{eq:gaussian_depth}
\end{equation}
where $\mathcal{R}_{\mathrm{depth}}(\cdot)$ denotes the Gaussian depth rendering operator, and $D_i^{3D}$ denotes the resulting Gaussian-derived depth. To better align the decoded Gaussian primitives with the underlying surface, we adopt the surface-aware parameterization of RaDe-GS~\cite{zhang2024radegs} instead of directly predicting conventional Gaussian primitives.

\subsection{Streaming and Patchwise Geometry Encoder}
\label{subsec:streaming_encoder}

The Streaming and Patchwise Geometry Encoder (SPGE) constructs the image tokens $\mathbf{F}_{i}$ used by the 2D stream of GSRT, targeting calibrated views with varying image resolution and sequence lengths. Its design contains two parts: (1) inspired by arbitrary-resolution geometry models \cite{cui2026anyresanygeo}, SPGE avoids resizing all views to a fixed size and instead processes each view as a set of spatially organized patches; (2) inspired by streaming geometry transformers \cite{lan2025stream3r,zhuo2026streamvggt}, SPGE processes views causally with cached historical tokens to support variable-length inputs.

\paragraph{Arbitrary-resolution Patch Encoding.}
For each calibrated view, we render a coarse Gaussian depth map $D_i^{0}$ and a normal map $N_i^{0}$ from the input 3DGS representation. These rendered cues provide metric geometry priors that complement the image. We then use DINOv2~\cite{oquab2023dinov2} to encode the image $I_i$ together with these rendered geometry cues, producing the image tokens $\mathbf{F}_{i}$ used in Eq.~\eqref{eq:gsrt_2d_stream}:
\begin{equation}
\mathbf{F}_i=\operatorname{DINOv2}(I_i,D_i^{0},N_i^{0}).
\end{equation}

The image tokens $\mathbf{F}_i$ are then divided into non-overlapping patches according to the original image size. Boundary patches are padded and masked when necessary, preserving the valid image region and aspect ratio. Following the multi-patch formulation of~\cite{cui2026anyresanygeo}, we assign each patch a global image coordinate and use alternating intra-patch and cross-patch attention to preserve local detail while propagating global layout information:
\begin{equation}
\bar{\mathbf{F}}_{i}
=
\operatorname{PatchEncoder}
\left(
\{\mathbf{F}_{i,m}\}_{m=1}^{M_i},
\{\mathbf{\mu}^{g}_{i,m}\}_{m=1}^{M_i}
\right),
\end{equation}
where $\mathbf{F}_{i,m}$ is the $m$-th patch token, $\mathbf{\mu}^{g}_{i,m}$ is its global image coordinate, $M_i$ is the number of valid patches determined by the image resolution of view $i$, and $\bar{\mathbf{F}}_{i}$ denotes the patch-encoded image tokens before multi-view aggregation.

\paragraph{Streaming View Aggregation.}
Processing all patches from all views in a single global sequence would lead to memory consumption that grows rapidly with both image resolution and the number of input views. Following the causal streaming design of STream3R and StreamVGGT~\cite{lan2025stream3r,zhuo2026streamvggt}, we process the $N$ views sequentially and reuse cached key--value (KV) representations from previous views. For view $i$, the patch-encoded tokens first undergo intra-view self-attention and then perform causal attention over the cached representations from previous views:

\begin{equation}
\bar{\mathbf{F}}_{i}
\leftarrow
\operatorname{CausalAttn}
\left(
\bar{\mathbf{F}}_{i},\mathcal{M}_{i-1}
\right),
\end{equation}
where $\mathcal{M}_{i-1}=\{\operatorname{KV}(\tilde{\mathbf{F}}_{j})\}_{j=1}^{i-1}$ contains the cached key--value pairs from previously processed views. The operator $\operatorname{CausalAttn}$ uses the current tokens as queries while attending only to historical context. During training, all $N$ views are processed simultaneously using a causal attention mask that prevents access to future views. During inference, $\mathcal{M}_{i}$ is updated incrementally by appending the current key--value pairs, thereby avoiding redundant computation for previously processed views. The resulting $\bar{\mathbf{F}}_{i}$ is then forwarded to GSRT as the image-token representation of view $i$.

\subsection{Scale-Aligned Hybrid Depth Refiner}
\label{subsec:depth_refiner}

The fused visual tokens produced by GSRT contain rich high-frequency cues, but the decoded visual depth $D_i^{2D}$ in Eq.~\eqref{eq:visual_depth} is still a monocular prediction and may be ambiguous up to a global scale and shift. In contrast, the Gaussian-rendered depth $D_i^{3D}$ in Eq.~\eqref{eq:gaussian_depth} is expressed in the calibrated coordinate frame of the input 3DGS scene, yet it can be over-smoothed around thin structures, occlusion boundaries, and sparsely covered regions. Inspired by recent high-resolution metric depth estimators that combine global scene context with fine local evidence~\cite{bochkovskii2025depthpro,li2024patchfusion}, we design SHDR to align the visual depth to the Gaussian metric frame and then fuse the two sources into a refined depth map.

Before fusion, a lightweight alignment head
$\psi_{\mathrm{align}}$, comprising a shallow convolutional encoder,
global pooling, and two linear predictors, maps the relative depth to
the Gaussian metric frame:
\begin{equation}
\begin{aligned}
(a_i,b_i)
&=\psi_{\mathrm{align}}(I_i,D_i^{3D},D_i^{2D}),\\
\tilde{D}_{i}^{2D}
&=a_iD_i^{2D}+b_i,\qquad a_i>0,
\end{aligned}
\end{equation}

A PatchFusion-style network~\cite{li2024patchfusion} predicts a per-pixel fusion weight and residual:
\begin{equation}
\begin{aligned}
(W_i,\Delta D_i)
&=\mathcal{F}_{\mathrm{fusion}}
\left(I_i,D_i^{3D},\tilde{D}_{i}^{2D},
\left|D_i^{3D}-\tilde{D}_{i}^{2D}\right|\right),\\
D_i^{*}
&=W_i\odot\tilde{D}_{i}^{2D}
 +(1-W_i)\odot D_i^{3D}
 +\Delta D_i .
\end{aligned}
\end{equation}
Here $W_i\in[0,1]$ selects the more reliable source per pixel, while $\Delta D_i$ corrects residual errors. Thus, $D_i^{3D}$ anchors the metric scale and $\tilde{D}_{i}^{2D}$ preserves image-aligned details, yielding globally consistent depth with sharp boundaries.

\subsection{Loss Function}
We supervise the refined depth map and its induced surface normals using
\begin{equation}
\mathcal{L}_{depth}
=
\frac{1}{N}\sum_i
\|D_i^{*}-D_i^{gt}\|_1,
\end{equation}
\begin{equation}
\mathcal{L}_{normal}
=
\frac{1}{N}\sum_i
\left(
1-\langle\mathbf{n}(D_i^{*}),\mathbf{n}_i^{gt}\rangle
\right),
\end{equation}
where $D_i^{gt}$ denotes the metric ground-truth depth of view $i$, $\mathbf{n}_i^{gt}$ is the corresponding ground-truth normal map, and $\mathbf{n}(D_i^{*})$ denotes the normal map computed from the refined depth.

To improve cross-view alignment, we adopt the multi-view consistency loss $\mathcal{L}_{cons}$ ~\cite{chen2024pgsr}. The total objective is
\begin{equation}
\mathcal{L}
=
\lambda_d\mathcal{L}_{depth}
+
\lambda_n\mathcal{L}_{normal}
+
\lambda_c\mathcal{L}_{cons},
\end{equation}
where $\lambda_d$, $\lambda_n$, and $\lambda_c$ are the weights for depth, normal, and multi-view consistency losses, respectively. 

\section{Experiments}

\begin{table*}[t]
\centering
\footnotesize
\caption{Quantitative comparison on DTU. Our method achieves the best overall reconstruction accuracy, demonstrating superior geometric fidelity and effectiveness over existing approaches}
\label{tab:dtu}
\setlength{\tabcolsep}{1.3mm}
\begin{tabular}{lccccccccccccccccc}
\toprule
Method & 24 & 37 & 40 & 55 & 63 & 65 & 69 & 83 & 97 & 105 & 106 & 110 & 114 & 118 & 122 & Mean & Time \\
\midrule
3DGS & 2.14 & 1.53 & 2.08 & 1.68 & 3.49 & 2.21 & 1.43 & 2.07 & 2.22 & 1.75 & 1.79 & 2.55 & 1.53 & 1.52 & 1.50 & 1.966 & 20m\\
3DGS+GS2Mesh & 0.59 & 0.79 & 0.70 & 0.38 & 0.78 & 1.00  & 0.69 & 1.25 & 0.96 & 0.59 & 0.50 & \underline{0.68} & \textbf{0.37} & \underline{0.50}& \textbf{0.46} & 0.683 & 20m(83s)\\

SteepGS+GS2Mesh & 0.61 & 0.79 & 0.74 & \underline{0.37} & 0.78 & 1.16 & 0.72 & 1.18 & 0.98 & 0.59 & \underline{0.49} & \underline{0.68} & 0.43 & 0.51 & 0.49 & 0.701 & 23m(83s)\\

EDGS+GS2Mesh & 0.58 & 0.87 & \textbf {0.45} & \underline{0.37} & 0.81 & 1.05 & 0.70 & \underline{1.15} & 0.97 & 0.59 & \underline{0.49} & \textbf{0.67} & 0.40 & \underline{0.50} & \underline{0.48} & 0.672 & 35m(83s)\\

3DGS+Ours & \underline{0.52} &\textbf{0.74} & 0.61 & 0.38 & \textbf{0.71} & 0.92 & \underline{0.64} & 1.30  & 1.00 & 0.55 &  \textbf{0.43} & 0.82 & \underline{0.38} & 0.55 & \underline{0.48} & \underline{0.668} & 20m(30s)\\

SteepGS+Ours & 0.56 & \underline{0.78} & 0.61 & 0.38 & 0.78 & \underline{0.90} & \underline{0.64} & \textbf{1.12} & \underline{0.89} & \underline{0.54} & 0.57 & 0.74 & \underline{0.38} & 0.55 & 0.52 & 0.671 & 23m(30s)\\

EDGS+Ours & \textbf {0.51} & \underline {0.78} & \underline {0.58} & \textbf {0.36} & \underline {0.73} & \textbf {0.88} & \textbf {0.62} & \underline{1.15} & \textbf{0.87} & \textbf {0.52} & 0.53 & 0.71 & \textbf {0.37} & \textbf{0.48}& 0.52 & \textbf{0.641} & 35m(30s)\\ 

\midrule
SuGaR & 1.47 & 1.33 & 1.13 & 0.61 & 2.25 & 1.71 & 1.15 & 1.63 & 1.62 & 1.07 & 0.79 & 2.45 & 0.98 & 0.88 & 0.79 & 1.324 & 65m\\
MILo & 0.43 & 0.74 & 0.34 & \underline{0.37} & 0.80 & 0.74 & 0.70 & 1.21 & 1.22 & 0.66 & 0.62 & 0.80 & 0.37 & 0.76 & 0.48 & 0.683 & 25m\\
RaDe-GS & 0.46 & 0.73 & 0.33 & 0.38 & 0.79 & 0.75 & 0.76 & 1.19 & 1.20 & 0.65 & 0.61 & 0.84 & 0.35 & 0.66 & 0.46 & 0.677 & 11m\\
QGS & 0.38 & 0.62 & 0.37 & 0.38 & \underline{0.75} & 0.55 & 0.51 & 1.12 & 0.68 & 0.61 & 0.46 & 0.58 & 0.35 & 0.41 & 0.40 & 0.545 & 48m\\
PGSR & 0.40 & 0.60 & 0.39 & \underline{0.37} & 0.78 & 0.59 & 0.53 & 1.18 & 0.73 & \underline{0.51} & 0.49 & 0.69 & 0.31 & \underline{0.37} & 0.38 & 0.555 & 30m \\
2DGS & 0.48 & 0.91 & 0.39 & 0.39 & 1.01 & 0.83 & 0.81 & 1.36 & 1.27 & 0.76 & 0.70 & 1.40 & 0.40 & 0.76 & 0.52 & 0.799 & 15m\\
GGGS & \underline{0.37} & \textbf{0.51} & \underline{0.27} & \textbf{0.31} & 0.81 & \underline{0.43} & \textbf{0.42} & \textbf{1.04} & \underline{0.64} & 0.52 & \textbf{0.31} & \underline{0.56} & \underline{0.30} & \textbf{0.31} & \textbf{0.33} & \underline{0.475} & 25m\\
2DGS+Ours & 0.45 & 0.84 & 0.40 & \underline{0.37} & 0.84 & 0.80 & 0.67 & 1.21 & 0.95 & 0.66 & 0.60 & 0.72 & 0.37 & 0.55 & 0.37 & 0.653 & 15m(30s)\\
GGGS+Ours & \textbf{0.33} & \underline{0.54} & \textbf{0.23} & \textbf{0.31} & \textbf{0.73} & \textbf{0.41} & \underline{0.45} & \underline{1.07} & \textbf{0.61} & \textbf{0.46} & \underline{0.35} & \textbf{0.52} & \textbf{0.29} & \underline{0.37} & \underline{0.35} & \textbf{0.468} & 25m(30s) \\
\bottomrule
\end{tabular}%
\end{table*}

\begin{figure*}[t]
    \centering
    \includegraphics[width=\textwidth]
    {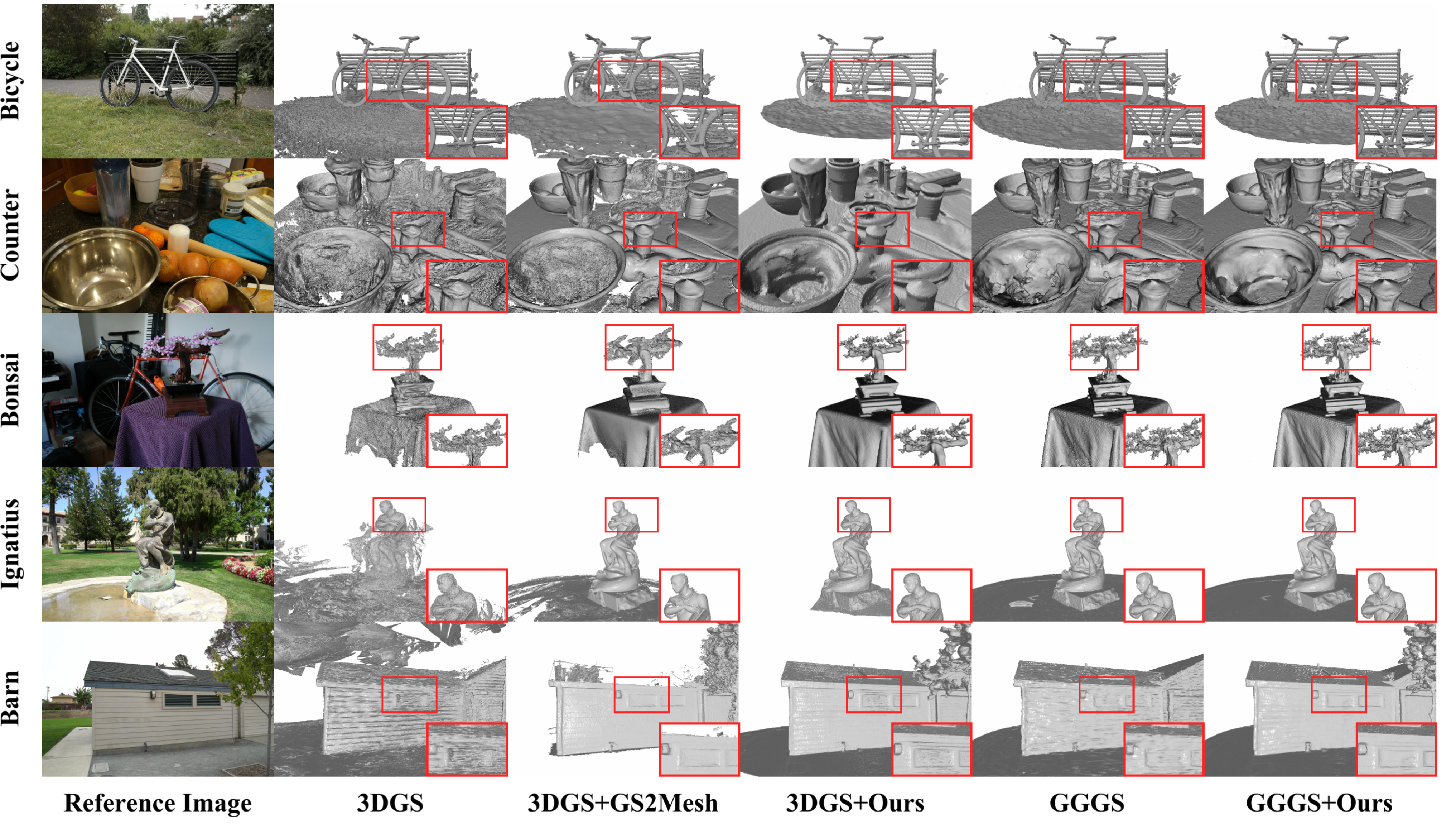}
    \caption{Qualitative comparison of mesh reconstruction results on Mip-NeRF 360 and Tanks and Temples.}
    \label{fig:qualitative}
\end{figure*}

\subsection{Experiment Setup}

\paragraph{Implementation Details.}
We implement AnyGS2Mesh in PyTorch and train the model on  NVIDIA A100 GPU. Each input scene consists of calibrated multi-view images and a corresponding 3DGS representation. For memory efficiency, we initialize the Gaussian primitives using PUP-3DGS~\cite{hanson2024pup3dgs} by filtering low-opacity and isolated primitives, retaining approximately 80K Gaussian primitives per scene. We optimize the model with AdamW for 30,000 iterations, using a weight decay of $1\times10^{-2}$. The learning rate is linearly warmed up for the first 1,000 iterations to a peak value of $2\times10^{-4}$ and then decayed using a cosine schedule. For network layers initialized from the pretrained VGGT weights, we apply a learning-rate multiplier of 0.1. Unless otherwise specified, we set $\lambda_d=1.0$, $\lambda_n=0.1$, and $\lambda_c=0.05$ in all experiments. For mesh extraction, we follow the standard volumetric fusion pipeline. Given the refined depth maps predicted for the calibrated views, we integrate them into a truncated signed distance field (TSDF)~\cite{curless1996volumetric} and extract the final mesh using Marching Cubes~\cite{lorensen1987marching}.

\paragraph{Datasets and Metrics.}

For training, we use a mixture of public datasets, including ARKitScenes~\cite{baruch2021arkitscenes}, Kubric~\cite{greff2022kubric}, BlendedMVS~\cite{yao2019blendedmvs}, ETH3D~\cite{schops2017eth3d}, CO3D-v2~\cite{reizenstein2021co3d}, Google Scanned Objects (GSO)~\cite{downs2022gso}, WildRGBD~\cite{xia2024wildrgbd}, Objaverse~\cite{deitke2023objaverse}, and the DTU~\cite{jensen2014dtu} training set. This diverse mixture spans synthetic and real-world data, indoor and outdoor scenes, and both object-centric and scene-level captures, providing broad geometric and appearance variations for learning feed-forward Gaussian-to-mesh reconstruction.

For evaluation, we report quantitative results on DTU~\cite{jensen2014dtu} and Tanks and Temples~\cite{knapitsch2017tanks}. On DTU, we evaluate the standard 15 scans and report Chamfer distance to the ground-truth point clouds. On Tanks and Temples, we report F-score on the four intermediate scenes (Barn, Caterpillar, Ignatius, and Truck). We further present qualitative results on Mip-NeRF 360~\cite{barron2022mipnerf360}. All methods are evaluated using the same scene splits and metrics.

\paragraph{Baselines.}
We compare against representative Gaussian-based methods from three groups. The rendering-oriented baseline EDGS \cite{kotovenko2026edgs} and SteepGS\cite{wang2025steepgs} provides strong novel-view synthesis but does not explicitly optimize mesh geometry. Reconstruction-oriented methods, including 2DGS~\cite{huang2024twodgs}, SuGaR~\cite{guedon2024sugar}, MILo~\cite{guedon2025milo}, RaDe-GS~\cite{zhang2024radegs}, QGS~\cite{zhang2025qgs}, PGSR~\cite{chen2024pgsr}, and GGGS~\cite{zhang2026geometry}, introduce surface-aware Gaussian parameterizations or geometric regularization for mesh extraction. We also compare with GS2Mesh~\cite{wolf2024gs2mesh}, the closest post-processing method for converting an existing Gaussian scene into a mesh. Since AnyGS2Mesh takes an existing Gaussian representation as input, it can be integrated as a plug-in module for different Gaussian pipelines. 

\subsection{Results and Evaluation}

\paragraph{Comparisons on DTU.}
Table~\ref{tab:dtu} reports the quantitative results on the DTU benchmark. When applied to the rendering-oriented 3DGS representation, AnyGS2Mesh achieves reconstruction accuracy comparable to GS2Mesh while reducing the post-processing time from 83 seconds to 30 seconds. Similar performance is observed when our method is integrated with EDGS and SteepGS, indicating that the proposed feed-forward reconstruction module is not tied to a particular Gaussian representation. More importantly, AnyGS2Mesh is complementary to surface-oriented Gaussian pipelines: it reduces the mean Chamfer distance of 2DGS from 0.799 to 0.653 and further improves GGGS from 0.475 to 0.468, yielding the best average result in Table 1. These results demonstrate that AnyGS2Mesh can efficiently recover accurate geometry from existing Gaussian representations without additional scene-specific mesh optimization.

\begin{table}[t]
\centering
\caption{Quantitative comparison on Tanks and Temples. Our method achieves the best overall reconstruction accuracy, obtaining the highest average score while introducing only marginal computational overhead}
\label{tab:tnt}
\footnotesize
\setlength{\tabcolsep}{3.2pt}
\begin{tabular}{lcccccc}
\toprule
Method & Barn & Cat. & Ign. & Truck & Mean  & Time\\
\midrule
3DGS & 0.13 & 0.08 & 0.04 & 0.19 & 0.110 &14m\\
3DGS+GS2Mesh & \underline{0.21} & \underline{0.17} & \textbf{0.64} & \underline{0.46} & \underline{0.370} &14m(196s) \\
3DGS+Ours & \textbf{0.35} &\textbf{0.26} & \underline{0.59}& \textbf{0.47} & \textbf{0.417} &  14m(45s)\\ \midrule
SuGaR & 0.14 & 0.16 & 0.33 & 0.26 & 0.223 &120m\\
RaDe-GS & 0.43 & 0.32 & 0.69 & 0.51 & 0.487 &18m\\
QGS & 0.55 & 0.40 & \textbf{0.81} & 0.64 & 0.600 &73m\\
PGSR & 0.52 & 0.38 & \underline{0.77} & 0.62 & 0.573 &45m\\
2DGS & 0.41 & 0.23 & 0.51 & 0.45 & 0.400 &34m\\
GGGS & \textbf{0.70} & \underline{0.56}& \textbf{0.81} & \underline{0.70} & \underline{0.693}& 32m\\
2DGS+Ours & 0.45 & 0.30 & 0.53 & 0.44 & 0.430 &34m(45s)\\
GGGS+Ours & \underline{0.68} & \textbf{0.59} & 0.76 & \textbf{0.78} & \textbf{0.703}& 32m(45s)\\
\bottomrule
\end{tabular}
\end{table}

\paragraph{Comparisons on Tanks and Temples.}
Table~\ref{tab:tnt} evaluates generalization on Tanks and Temples, which contains larger scenes and more incomplete observations than DTU. Starting from 3DGS, AnyGS2Mesh improves the mean F-score from 0.110 to 0.417 and outperforms GS2Mesh with less additional inference time. Although the standalone model remains behind strong surface-optimized methods under severe occlusion, it consistently improves existing pipelines: 2DGS increases from 0.400 to 0.430 and GGGS from 0.693 to 0.703, with GGGS+Ours achieving the best mean result. These results demonstrate the complementarity of our module with both rendering- and surface-oriented Gaussian representations.

\paragraph{Comparisons on Mip-NeRF 360.}
Mip-NeRF 360~\cite{barron2022mipnerf360} is used for qualitative evaluation on unbounded real-world scenes with wide camera baselines, complex backgrounds, and large spatial extents. As shown in Fig.~\ref{fig:qualitative}, AnyGS2Mesh can convert such Gaussian scenes into coherent meshes without resizing the input views to a fixed resolution or performing scene-specific mesh optimization. The qualitative results indicate that the streaming patchwise encoder helps preserve local structures in high-resolution views, while Gaussian-Guided reasoning provides metric geometric priors for broader scene regions. We therefore use Mip-NeRF 360 mainly to demonstrate cross-scene robustness and visual reconstruction behavior rather than reporting quantitative mesh metrics.

\subsection{Ablation Study}
\begin{table}[t]
\centering
\caption{Ablation study on DTU and Tanks and Temples.}
\label{tab:ablation}
\footnotesize

\begin{tabular*}{\linewidth}{
    @{\extracolsep{\fill}}
    lcc
    @{}
}
\toprule
\textbf{Variant} &
\textbf{TnT F1 $\uparrow$} &
\textbf{DTU Cf. $\downarrow$} \\
\midrule
Full model                  & \textbf{0.417} & \textbf{0.668} \\
\midrule
w/o 2D Visual Stream        & 0.108 & 1.830 \\
w/o 3D Gaussian Stream      & 0.282 & 0.882 \\
w/o Spatial Reasoning       & 0.346 & 0.791 \\
w/o SHDR                    & 0.315 & 0.921 \\
\bottomrule
\end{tabular*}
\end{table}

In this section, we evaluate the contribution of each component in AnyGS2Mesh. Table 3 reports the TnT F1 score and DTU Cf. (Chamfer distance). The complete AnyGS2Mesh configuration (Full model) achieves a TnT F1 score of 0.417 and a DTU Chamfer distance of 0.668. Removing the 2D visual stream (w/o 2D Visual Stream) causes the largest performance degradation, decreasing the TnT F1 score to 0.108 and increasing the DTU Chamfer distance to 1.830, which demonstrates the importance of high-resolution visual evidence. Removing the 3D Gaussian stream (w/o 3D Gaussian Stream) decreases the TnT F1 score to 0.282 and increases the DTU Chamfer distance to 0.882, confirming the importance of explicit metric Gaussian priors. Removing the spatial reasoning module (w/o Spatial Reasoning) results in a TnT F1 score of 0.346 and a DTU Chamfer distance of 0.791, showing that view-aligned cross-stream interaction is essential for effective feature fusion. Finally, removing the scale-aligned hybrid depth refinement module (w/o SHDR) decreases the TnT F1 score to 0.315 and increases the DTU Chamfer distance to 0.921, confirming the effectiveness of scale-aligned depth fusion. Overall, all components consistently contribute to reconstruction quality across both datasets.

\FloatBarrier
\section{Conclusion}
We presented AnyGS2Mesh, a feed-forward framework that converts existing 3D Gaussian Splatting representations into explicit meshes without scene-specific mesh optimization. By combining Gaussian-Guided spatial reasoning, native-resolution streaming visual encoding, and scale-aligned hybrid depth refinement, AnyGS2Mesh can serve as a plug-in reconstruction module for different Gaussian pipelines. Experiments on DTU and Tanks and Temples show that it improves existing Gaussian representations such as 2DGS and GGGS, while qualitative results on Mip-NeRF 360 demonstrate stable reconstruction on unbounded real-world scenes.

\FloatBarrier
\appendix
\setcounter{secnumdepth}{2}
\section*{Supplementary Material}
\section{Method and Network Architecture Details}
\label{sec:supp_method}

This section details two key components of \method{}: view-aligned 2D–3D fusion in GSRT and native-resolution streaming processing in SPGE.

\subsection{View-Aligned Spatial Reasoning in GSRT}
\label{subsec:supp_gsrt}

GSRT maintains a visual stream
$\mathbf{F}^{2\mathrm{D}}=\{\mathbf{f}^{2\mathrm{D}}_{i,m}\}$ and a sparse
Gaussian stream
$\mathbf{F}^{3\mathrm{D}}=\{\mathbf{f}^{3\mathrm{D}}_k\}$. Here $i$ indexes
the input view, $m$ indexes a visual patch in its native image grid, and $k$
indexes an occupied Gaussian voxel. The two streams use the same hidden width
$C$ after separate linear input projections. Spatial interaction is performed
only between a visual patch and the Gaussian tokens that project to its image
support, avoiding dense attention between all 2D and 3D tokens.

\paragraph{Projection and patch assignment.}
Let $\mathbf{p}_k\in\mathbb{R}^3$ denote the representative world-space
position of Gaussian token $k$. Its camera-space position and image projection
in view $i$ are
\begin{equation}
\begin{aligned}
\mathbf{x}_{ik}
&= \mathbf{R}_i\mathbf{p}_k+\mathbf{t}_i,\\
\mathbf{u}_{ik}
&= \pi\!\left(\mathbf{K}_i\mathbf{x}_{ik}\right).
\end{aligned}
\label{eq:supp_gsrt_projection}
\end{equation}
Here,
$\pi([x,y,z]^\top)=[x/z,y/z]^\top$. We write
$z_{ik}=[\mathbf{x}_{ik}]_z$ for the camera-space depth. For a patch support
of $p_h\times p_w$ pixels, the patch containing a projected point
$\mathbf{u}=(u_x,u_y)$ is
\begin{equation}
\begin{aligned}
m_i(\mathbf{u})
&= \left\lfloor\frac{u_y}{p_h}\right\rfloor P_i^w
 + \left\lfloor\frac{u_x}{p_w}\right\rfloor,\\
P_i^w
&= \left\lceil\frac{W_i}{p_w}\right\rceil.
\end{aligned}
\label{eq:supp_gsrt_patch_index}
\end{equation}
Only tokens with positive camera-space depth and projections inside the valid,
unpadded image region are retained.

\paragraph{Visibility and neighborhood construction.}
We use the coarse depth $D_i^0$ rendered from the input
3DGS~\cite{kerbl20233dgs} to reject tokens
that are geometrically inconsistent with the first visible surface. A token is
marked visible when
\begin{equation}
\operatorname{vis}(i,k)
=
\mathbf{1}\!\left[
\begin{gathered}
z_{ik}>0,\qquad
\mathbf{u}_{ik}\in\Omega_i,\\
\frac{
\left|z_{ik}-D_i^0(\mathbf{u}_{ik})\right|
}{
D_i^0(\mathbf{u}_{ik})+\epsilon
}
\leq \tau_{\mathrm{vis}}
\end{gathered}
\right].
\label{eq:supp_gsrt_visibility}
\end{equation}
Here, $\mathbf{1}[\cdot]$ denotes the indicator function,
$\Omega_i$ is the valid native-resolution image domain, and $\epsilon$ is a
small constant introduced for numerical stability. We set
$\tau_{\mathrm{vis}}=0.1$ in all experiments. Therefore, a Gaussian token is
considered visible only if (i) it lies in front of the camera
($z_{ik}>0$), (ii) its projected center lies inside the valid image domain,
and (iii) its camera-space depth differs from the rendered surface depth by
no more than $10\%$ in relative terms. The rendered depth is bilinearly
sampled at non-integer projections, and pixels without valid rendered depth are
excluded from the association.

The candidate neighborhood of patch $m$ is consequently
\begin{equation}
\mathcal{N}_{i,m}
= \left\{
k\,\middle|\,
m_i(\mathbf{u}_{ik})=m,
\;\operatorname{vis}(i,k)=1
\right\}.
\label{eq:supp_gsrt_neighborhood}
\end{equation}
To bound computation, at most \(K_g=50\) candidates are retained for each patch.
If more than 50 candidates satisfy the visibility criterion, they are ranked
in ascending order of the relative depth residual in
Eq.~\eqref{eq:supp_gsrt_visibility}, and the 50 candidates with the smallest
residuals are retained. If \(\mathcal{N}_{i,m}\) is empty, the Gaussian feature
associated with patch m is set to a zero vector. 

\paragraph{Patch-aligned 3DGS feature construction.}
The projection, visibility filtering, and neighborhood selection described
above are performed before the features are passed to the visual aggregator.
After selection and zero padding, the patch-aligned 3DGS features are arranged
as a fixed-size tensor
\begin{equation}
\mathbf{P}
\in
\mathbb{R}^{B\times S\times H_p\times W_p\times K_g\times C_g},
\qquad
C_g=96,
\quad K_g=50,
\label{eq:supp_gsrt_pointfeat_tensor}
\end{equation}
where $B$ is the batch size, $S$ is the number of input views, and
$H_p\times W_p$ is the visual patch grid. For a flattened patch index $m$,
let $\mathbf{p}_{i,m,n}\in\mathbb{R}^{C_g}$ denote the $n$-th patch-aligned
3DGS feature, where $n\in\{1,\ldots,K_g\}$. A shared linear layer projects
each feature to the visual hidden width:
\begin{equation}
\mathbf{g}_{i,m,n}
= \mathbf{W}_{g}\mathbf{p}_{i,m,n}+\mathbf{b}_{g},
\qquad
\mathbf{g}_{i,m,n}\in\mathbb{R}^{C}.
\label{eq:supp_gsrt_point_projection}
\end{equation}
The projected features belonging to the same patch are stacked to form its
local 3DGS context,
\begin{equation}
\mathbf{G}_{i,m}
=
\left[
\mathbf{g}_{i,m,1},
\ldots,
\mathbf{g}_{i,m,K_g}
\right]^{\top}
\in
\mathbb{R}^{K_g\times C}.
\label{eq:supp_gsrt_patch_context}
\end{equation}

\paragraph{Patch-wise 3DGS-to-2D cross-attention.}
Before 3DGS feature fusion, the image patch tokens are augmented with the
embedded depth and normal inputs . Let
$\mathbf{f}^{2\mathrm{D}}_{i,m}\in\mathbb{R}^{C}$ denote the resulting visual
token of patch $m$ in view $i$. The implementation reshapes each patch token
as a single-query sequence and supplies its patch-aligned 3DGS features as the
local context:
\begin{equation}
\mathbf{h}_{i,m}
= \operatorname{CrossAttnBlock}\!\left(
\mathbf{f}^{2\mathrm{D}}_{i,m},
\mathbf{G}_{i,m}
\right),
\qquad
\mathbf{h}_{i,m}\in\mathbb{R}^{C}.
\label{eq:supp_gsrt_cross_attention}
\end{equation}
Consequently, a visual patch token attends only to the 3DGS features assigned
to the same patch, rather than to all Gaussian tokens in the scene. The
cross-attention block produces one 3DGS-conditioned token for every visual
patch. In the current architecture, this fusion operation is applied once after the
final  attention group.

\paragraph{Patch-level refinement.}
For view $i$, collect the original visual patch tokens and their
3DGS-conditioned counterparts as
\begin{equation}
\begin{aligned}
\mathbf{F}^{2\mathrm{D}}_i
&=
\left[
\mathbf{f}^{2\mathrm{D}}_{i,1},
\ldots,
\mathbf{f}^{2\mathrm{D}}_{i,M}
\right]^{\top},\\
\mathbf{H}_i
&=
\left[
\mathbf{h}_{i,1},
\ldots,
\mathbf{h}_{i,M}
\right]^{\top},
\qquad M=H_pW_p.
\end{aligned}
\label{eq:supp_gsrt_patch_sequences}
\end{equation}
The two sequences are concatenated along the token dimension,
\begin{equation}
\mathbf{Z}_i
= \operatorname{concat}\!\left(
\mathbf{F}^{2\mathrm{D}}_i,
\mathbf{H}_i
\right)
\in
\mathbb{R}^{2M\times C},
\label{eq:supp_gsrt_patch_concat}
\end{equation}
and refined using a patch-level attention layer:
\begin{equation}
\widetilde{\mathbf{Z}}_i
= \operatorname{Attn}_{\mathrm{patch}}\!\left(
\mathbf{Z}_i
\right).
\label{eq:supp_gsrt_patch_attention}
\end{equation}
Only the first $M$ output positions, which correspond to the original visual
patch-token positions, are retained:
\begin{equation}
\widehat{\mathbf{F}}^{2\mathrm{D}}_i
= \widetilde{\mathbf{Z}}_i[1\!:\!M].
\label{eq:supp_gsrt_patch_refinement}
\end{equation}
The camera token and register tokens are excluded from the 3DGS fusion module
and concatenated back with the refined patch tokens without modification.

\subsection{SPGE for Native-Resolution and Streaming Inputs}
\label{subsec:supp_spge}

SPGE preserves the aspect ratio and valid image extent of each observation.
It consists of a geometry-conditioned visual front end, native-resolution
patch packing, and causal view aggregation. No input view is resized to a
shared spatial shape.

\paragraph{Geometry-conditioned visual tokens.}
For each view $i$, we use the RGB image $\mathbf{I}_i$, an initial metric
depth map $D_i^0$, a normal map $\mathbf{N}_i^0$, and a valid geometry mask
$\mathbf{M}_i^0$. The RGB
image is first encoded by the pretrained DINOv2 patch
stem~\cite{oquab2023dinov2}, while depth and
normal are encoded by two lightweight patch embedding adapters with the same
patch size:
\begin{align}
\mathbf{F}^{\mathrm{rgb}}_i
&=
E_{\mathrm{DINO}}(\mathbf{I}_i),
\label{eq:supp_spge_rgb}
\\
\bar{D}_i^0
&=
\operatorname{NormDepth}(D_i^0,\mathbf{M}_i^0),
\label{eq:supp_spge_depth_norm}
\\
\mathbf{F}^{\mathrm{dep}}_i
&=
E_{\mathrm{dep}}
\left(
[\bar{D}_i^0,\mathbf{M}_i^0]
\right),
\label{eq:supp_spge_depth}
\\
\mathbf{F}^{\mathrm{nor}}_i
&=
E_{\mathrm{nor}}
\left(
[\mathbf{N}_i^0,\mathbf{M}_i^0]
\right).
\label{eq:supp_spge_normal}
\end{align}
Here, $E_{\mathrm{dep}}$ is a patch embedding layer with two input channels
for normalized depth and mask, and $E_{\mathrm{nor}}$ is a patch embedding
layer with four input channels for normal and mask. The depth normalization is
computed only over valid pixels:
\begin{equation}
\bar{D}_i^0(\mathbf{u})
=
\frac{D_i^0(\mathbf{u})}
{
\operatorname{mean}_{\mathbf{v}:\mathbf{M}_i^0(\mathbf{v})=1}
D_i^0(\mathbf{v}) + \epsilon
}
\cdot
\mathbf{M}_i^0(\mathbf{u}).
\label{eq:supp_spge_depth_normalization}
\end{equation}

The geometry-conditioned patch token is then obtained by additive token-level
fusion:
\begin{equation}
\mathbf{F}_i
=
\mathbf{F}^{\mathrm{rgb}}_i
+
\mathbf{F}^{\mathrm{dep}}_i
+
\mathbf{F}^{\mathrm{nor}}_i .
\label{eq:supp_spge_token_fusion}
\end{equation}
This keeps the pretrained RGB encoder unchanged and injects geometry through
separate depth and normal token adapters.

\paragraph{Camera-conditioned token sequence.}
In addition to patch-level geometry conditioning, AnyGS2Mesh injects camera
information through a special camera token. Given the camera extrinsic
$\mathbf{T}_i$ and intrinsic $\mathbf{K}_i$, a pose encoding function
$\psi_{\mathrm{cam}}$ maps them to a camera feature:
\begin{equation}
\mathbf{p}_i
=
\psi_{\mathrm{cam}}(\mathbf{T}_i,\mathbf{K}_i),
\qquad
\mathbf{c}_i
=
\mathbf{c}_0
+
\mathbf{W}_{\mathrm{cam}}\mathbf{p}_i ,
\label{eq:supp_spge_camera_token}
\end{equation}
where $\mathbf{c}_0$ is a learnable camera token. The final input sequence for
view $i$ is
\begin{equation}
\mathbf{S}_i
=
\operatorname{concat}
\left(
\mathbf{c}_i,
\mathbf{R},
\mathbf{F}_i
\right),
\label{eq:supp_spge_final_sequence}
\end{equation}
where $\mathbf{R}$ denotes learnable register tokens and $\mathbf{F}_i$
contains all geometry-conditioned image patch tokens.

\paragraph{Native-resolution packing and coordinates.}
Let the valid size of view $i$ be $H_i\times W_i$ and the patch stride be
$p_h\times p_w$. We preserve the native aspect ratio and pad only the bottom
and right boundaries, giving
\begin{equation}
P_i^h=\left\lceil\frac{H_i}{p_h}\right\rceil,
\qquad
P_i^w=\left\lceil\frac{W_i}{p_w}\right\rceil,
\qquad
M_i=P_i^hP_i^w.
\label{eq:supp_spge_patch_count}
\end{equation}
Each patch token is paired with a validity mask $\mathbf A_i$ and a coordinate
in the global patch grid. For a patch centered at
$(c^x_{i,m},c^y_{i,m})$, we define
\begin{equation}
\mathbf p^g_{i,m}
=
\left(
\frac{c^x_{i,m}}{p_w},
\frac{c^y_{i,m}}{p_h}
\right).
\label{eq:supp_spge_global_coordinate}
\end{equation}
The coordinate is never reset when a view is divided into local windows, so
tokens from different windows remain aligned in the native image coordinate
system. Following the global-position design of Any Resolution Any
Geometry~\cite{cui2026anyresanygeo},
we apply 2D rotary positional encoding~\cite{su2021roformer} to the query and
key vectors:
\begin{equation}
\begin{aligned}
\widetilde{\mathbf q}_{i,m}
&=\operatorname{RoPE}\!\left(
\mathbf q_{i,m},\mathbf p^g_{i,m}
\right),\\
\widetilde{\mathbf k}_{i,m}
&=\operatorname{RoPE}\!\left(
\mathbf k_{i,m},\mathbf p^g_{i,m}
\right).
\end{aligned}
\label{eq:supp_spge_global_rope}
\end{equation}
For batching, variable-length view sequences are concatenated with offsets
$o_1=0$ and $o_i=\sum_{j<i}M_j$. Camera and register tokens retain their
dedicated embeddings, while padded patch positions are excluded by
$\mathbf A_i$.

\paragraph{Local and cross-patch encoding.}
Inspired by the intra- and cross-patch reasoning in Any Resolution Any
Geometry, SPGE alternates local attention within spatial windows and
cross-patch communication across windows:
\begin{align}
\mathbf Z_i^{\ell+1}
&=\Phi_{\mathrm{local}}\!\left(
\mathbf Z_i^{\ell},\mathbf p_i^g,\mathbf A_i
\right),
\label{eq:supp_spge_local}
\\
\mathbf Z_i^{\ell+2}
&=\Phi_{\mathrm{cross}}\!\left(
\mathbf Z_i^{\ell+1},\mathbf p_i^g,\mathbf A_i
\right).
\label{eq:supp_spge_cross}
\end{align}
The local block preserves fine structures within each window, whereas the
cross-patch block exchanges information between spatially separated regions.
Global RoPE keeps these interactions consistent across window boundaries and
reduces patch-seam artifacts without resizing the input image.

\paragraph{Causal view aggregation.}
After patch encoding, the views are aggregated in input
order~\cite{lan2025stream3r,zhuo2026streamvggt}. Let
$\overline{\mathbf F}_i^{\ell}$ denote the tokens of view $i$ at streaming
layer $\ell$. The current view provides the queries, while the keys and values
include both the current tokens and a cache of preceding views:
\begin{equation}
\begin{aligned}
\mathbf Q_i^{\ell}
&=\mathbf W_Q^{\ell}\overline{\mathbf F}_i^{\ell},\\
\mathbf K_{\leq i}^{\ell}
&=\left[
\mathcal C_{i-1}^{\ell,K};
\mathbf W_K^{\ell}\overline{\mathbf F}_i^{\ell}
\right],\\
\mathbf V_{\leq i}^{\ell}
&=\left[
\mathcal C_{i-1}^{\ell,V};
\mathbf W_V^{\ell}\overline{\mathbf F}_i^{\ell}
\right].
\end{aligned}
\label{eq:supp_spge_qkv}
\end{equation}
The streaming update and cache update are
\begin{align}
\widetilde{\mathbf F}_i^{\ell}
&=\overline{\mathbf F}_i^{\ell}
+\operatorname{Attn}\!\left(
\mathbf Q_i^{\ell},
\mathbf K_{\leq i}^{\ell},
\mathbf V_{\leq i}^{\ell}
\right),
\label{eq:supp_spge_causal_attn}
\\
\mathcal C_i^{\ell}
&=\operatorname{Append}\!\left(
\mathcal C_{i-1}^{\ell},
\operatorname{KV}^{\ell}\!\left(
\widetilde{\mathbf F}_i^{\ell}
\right)
\right).
\label{eq:supp_spge_cache_update}
\end{align}

\section{Training Data and Implementation Details}
\label{sec:supp_training}

In this section, we provide additional details of our training protocol, model initialization, and experiments.

\subsection{Training datasets.}
We train the model on a heterogeneous mixture of nine public datasets:
ARKitScenes~\cite{baruch2021arkitscenes},
Kubric~\cite{greff2022kubric}, BlendedMVS~\cite{yao2019blendedmvs},
ETH3D~\cite{schops2017eth3d}, CO3D-v2~\cite{reizenstein2021co3d},
Google Scanned Objects~\cite{downs2022gso},
WildRGBD~\cite{xia2024wildrgbd}, Objaverse~\cite{deitke2023objaverse}, and
the DTU training split~\cite{jensen2014dtu}. These datasets cover
synthetic and real data, indoor and outdoor environments, and both
object-centric and scene-level captures. This diversity exposes the model to
substantial variation in scene geometry, appearance, camera motion, object
scale, and reconstruction quality, thereby improving convergence and
generalization across different scene types.

All nine datasets provide usable metric depth ground truth, either directly
from RGB-D sensors or synthetic rendering, or indirectly from registered
scans, reconstructed point clouds, or watertight meshes from which depth can
be rendered. Surface-normal supervision is taken from the original dataset
when available; otherwise, it is computed from the metric depth maps or the
corresponding 3D geometry. The calibrated RGB images and depth maps can
therefore be organized into multi-view scene samples and used for
geometry-aware scene training. They also provide the geometric observations
required to construct the input 3DGS representation.

During training, one dataset is sampled at each iteration according to a
dataset-level probability distribution. The number of retained scenes or
samples and the final sampling probabilities will be reported together with the released training
configuration. Table~\ref{tab:supp_dataset_breakdown} summarizes the dataset
types and their sources of depth supervision.

\begin{table*}[t]
    \centering
    \caption{Training-data composition and dataset-level sampling
distribution used in our experiments. Approximately 100--200 scenes,
sequences, or objects are selected from each dataset when available.
For datasets containing fewer scenes, all available training scenes
are used.}
    \label{tab:supp_dataset_breakdown}
    \scriptsize
    \setlength{\tabcolsep}{6pt}
    \renewcommand{\arraystretch}{1.15}
    \begin{tabular}{
        @{}
        p{3.2cm}
        p{3.0cm}
        p{2.2cm}
        p{3.3cm}
        c
        @{}
    }
        \toprule
        Dataset
        & Scene type
        & Domain
        & Selected scenes/samples
        & Sampling probability \\
        \midrule

        ARKitScenes
        & Indoor scenes
        & Real
        & 200 scenes
        & 18\% \\

        Kubric
        & Mixed scenes
        & Synthetic
        & 200 generated scenes
        & 12\% \\

        BlendedMVS
        & Indoor/outdoor
        & Real-derived
        & 113 scenes
        & 12\% \\

        ETH3D
        & Indoor/outdoor
        & Real
        & 13 scenes (all)
        & 6\% \\

        CO3D-v2
        & Object-centric
        & Real
        & 200 object sequences
        & 14\% \\

        Google Scanned Objects
        & Object-centric
        & Real scans
        & 150 objects
        & 8\% \\

        WildRGBD
        & Object-centric
        & Real
        & 200 object sequences
        & 12\% \\

        Objaverse
        & Object-centric
        & Synthetic
        & 200 objects
        & 10\% \\

        DTU training split
        & Object/indoor
        & Real
        & 113 scenes
        & 8\% \\

        \bottomrule
    \end{tabular}
\end{table*}

\subsection{Construction of Training Inputs}
\label{subsec:supp_input_construction}

Given a scene with multi-view images, we independently construct its
scene-level 3DGS representation and generate the corresponding view-wise
training inputs following
Algorithm~\ref{alg:supp_training_input_construction}. If a scene contains
more than 300 images, we randomly sample 300 views; otherwise, all available
views are retained.

\begin{algorithm}[t]
    \caption{Construction of Scene-Level 3DGS Training Inputs}
    \label{alg:supp_training_input_construction}
    \begin{algorithmic}[1]
        \Require Scene $s$ with multi-view images
        $\mathcal{I}_s=\{\mathbf{I}_i\}_{i=1}^{V_s}$;
        optional camera calibration $\mathcal{C}_s$;
        optional scene geometry $\mathcal{P}_s$;
        target Gaussian count $N_g=80{,}000$
        \Ensure Processed 3DGS $\mathcal{G}_s$ and view-wise inputs
        $\{\mathcal{X}_i\}_{i=1}^{V_s}$

        \If{camera intrinsics and poses are provided}
            \State $\{\mathbf{K}_i,\mathbf{T}_i\}_{i=1}^{V_s}
            \gets \mathcal{C}_s$
            \State $\mathcal{P}_s^0
            \gets \Call{InitializeGeometry}{
                \mathcal{I}_s,\mathcal{C}_s,\mathcal{P}_s
            }$
        \Else
            \State $\{\mathbf{K}_i,\mathbf{T}_i\}_{i=1}^{V_s},
            \mathcal{P}_s^0
            \gets \Call{COLMAP}{\mathcal{I}_s}$~\cite{schoenberger2016sfm}
            \If{camera registration or sparse reconstruction fails}
                \State \Return \textsc{InvalidScene}
            \EndIf
        \EndIf

        \State $\mathcal{G}_s^0
        \gets \Call{Initialize3DGS}{\mathcal{P}_s^0}$

        \State $\mathcal{G}_s
        \gets \Call{Train3DGS}{
            \mathcal{G}_s^0,
            \mathcal{I}_s,
            \{\mathbf{K}_i,\mathbf{T}_i\}_{i=1}^{V_s},
            30{,}000
        }$
        \Comment{Original 3DGS optimization~\cite{kerbl20233dgs}}

        \If{$|\mathcal{G}_s|>N_g$}
            \State $\mathcal{G}_s
            \gets \Call{PUPPrune}{\mathcal{G}_s,N_g}$
            \Comment{Prune to approximately 80K Gaussians~\cite{hanson2024pup3dgs}}
        \EndIf

        \For{$i=1,\ldots,V_s$}
            \State $(\mathbf{I}_i^{\mathrm{gs}},
            D_i^{\mathrm{gs}},
            \mathbf{N}_i^{\mathrm{gs}},
            \mathbf{M}_i^{\mathrm{gs}})
            \gets
            \Call{Render}{
                \mathcal{G}_s,\mathbf{K}_i,\mathbf{T}_i
            }$

            \If{\Call{GeometryReliable}{
                $D_i^{\mathrm{gs}},\mathbf{M}_i^{\mathrm{gs}}$
            }}
                \State $D_i^0 \gets D_i^{\mathrm{gs}}$
                \State $\mathbf{M}_i^0
                \gets \mathbf{M}_i^{\mathrm{gs}}$
            \Else
                \State $\mathcal{V}_i
                \gets \Call{SelectSourceViews}{
                    i,\{\mathbf{K}_j,\mathbf{T}_j\}_{j=1}^{V_s}
                }$

                \State $D_i^0
                \gets \Call{DLNR}{
                    \mathbf{I}_i,\mathbf{K}_i,\mathbf{T}_i,
                    \{\mathbf{I}_j,\mathbf{K}_j,\mathbf{T}_j
                    \mid j\in\mathcal{V}_i\}
                }$
                \Comment{Replace only the unreliable depth~\cite{wolf2024gs2mesh}}

                \State $\mathbf{M}_i^0
                \gets \Call{ValidDepthMask}{D_i^0}$
            \EndIf

            \State $\mathcal{X}_i
            \gets
            \{
                \mathbf{I}_i^{\mathrm{gs}},
                D_i^0,
                \mathbf{N}_i^{\mathrm{gs}},
                \mathbf{M}_i^0,
                \mathbf{K}_i,
                \mathbf{T}_i,
                \mathcal{G}_s
            \}$
        \EndFor

        \State \Return
        $\mathcal{G}_s,\{\mathcal{X}_i\}_{i=1}^{V_s}$
    \end{algorithmic}
\end{algorithm}
\subsection{Model Initialization}
\label{subsec:supp_evaluation}
To leverage pretrained multi-view geometric priors, we initialize the
VGGT-compatible visual backbone with pretrained VGGT
weights~\cite{tang2023vggt} and fine-tune it
together with the proposed modules. The input scene-level 3DGS representation
and camera parameters remain fixed throughout training, while PUP-3DGS pruning
and optional DLNR depth estimation are performed offline. The visual backbone,
SPGE encoder, PTV3-based Gaussian stream~\cite{wu2024pointtransformerv3},
Visual--Gaussian spatial reasoning
layers, depth and Gaussian prediction heads, and SHDR modules are jointly
optimized. The Gaussian renderer, TSDF fusion~\cite{curless1996volumetric},
and Marching Cubes~\cite{lorensen1987marching} contain no learnable
parameters.

\FloatBarrier
\section{More Results}
\label{sec:supp_results}

This section provides additional experimental results from four aspects:
the runtime scalability of SPGE with different numbers of input views,
the compatibility of \method{} as a post-processing module, qualitative
comparisons with existing Gaussian-to-mesh methods, and an analysis of
the intermediate depth and geometry predictions produced by different
model components.

\subsection{View-Count Scalability and Native-Resolution Processing}
\label{subsec:supp_view_runtime}

\begin{figure}[t]
    \centering
    \includegraphics[width=\linewidth]
    {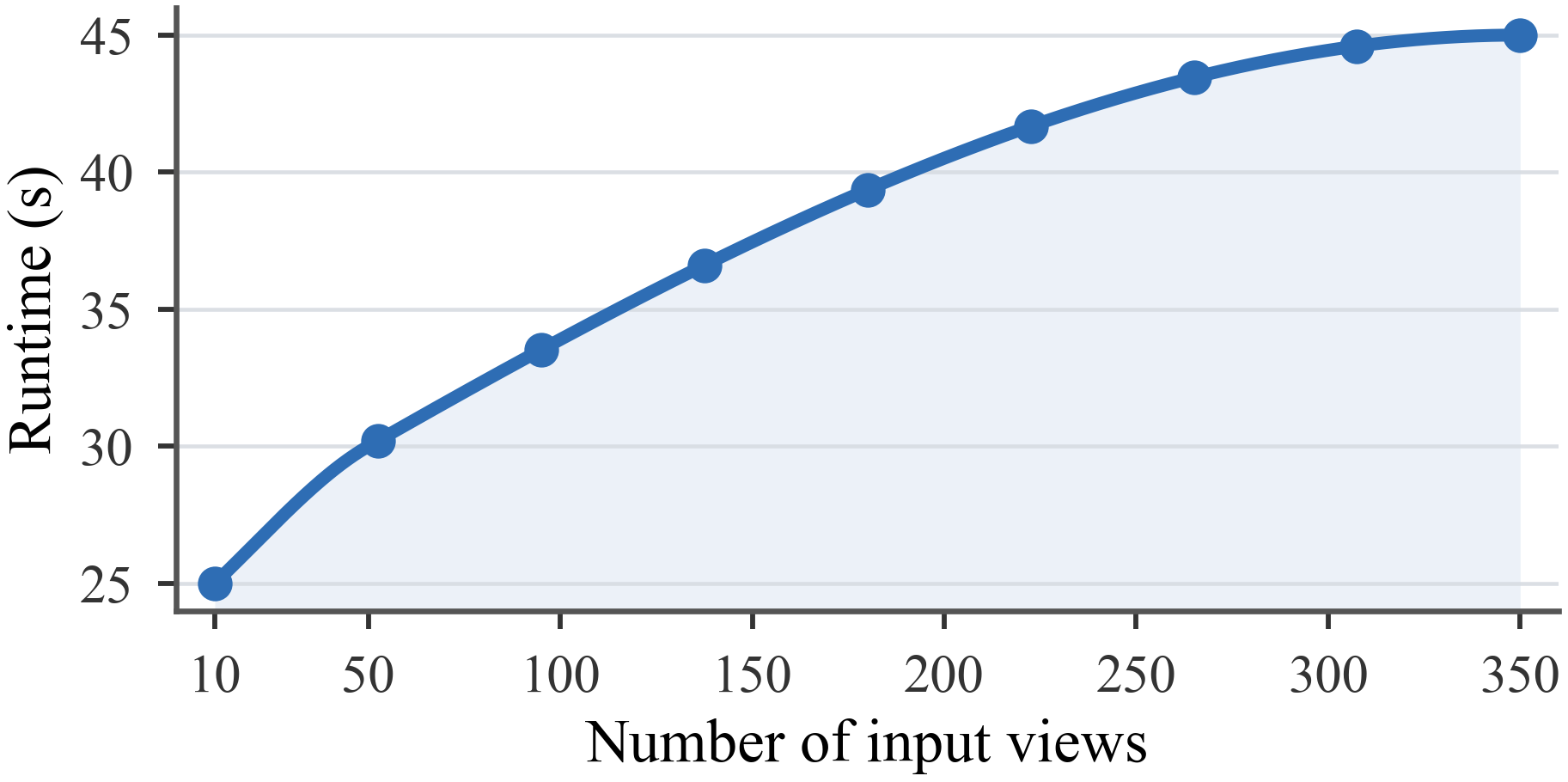}
    \caption{Runtime of SPGE with different numbers of input views.
    The runtime increases gradually as more views are introduced, from
    approximately 25 seconds for 10 views to 45 seconds for 350 views.}
    \label{fig:supp_view_runtime}
\end{figure}

Figure~\ref{fig:supp_view_runtime} analyzes the computational scalability of
SPGE with respect to the number of input views. As expected, its runtime
increases with the view count because each additional image introduces new
patch tokens and requires an additional causal aggregation step. Nevertheless,
the increase is gradual: the runtime grows from approximately 25 seconds for
10 views to only 45 seconds for 350 views. This result demonstrates that SPGE
can efficiently exploit substantially larger view sets without imposing a
fixed architectural upper bound on the number of input images.

More importantly, this additional computation provides flexibility that is
particularly valuable for large-scale scene reconstruction. SPGE preserves
the native resolution and aspect ratio of every observation and processes
variable-length view sequences through patchwise encoding and cached causal
aggregation. Consequently, it can accept an arbitrary number of views with
different resolutions, subject only to available computational resources,
rather than requiring all images to be resized to a shared spatial shape.
Increasing the number of observations also provides broader scene coverage
and more overlapping geometric evidence, allowing the model to establish
stronger cross-view correspondences and reason about scene-level geometric
consistency.

In contrast, the standard VGGT formulation~\cite{tang2023vggt} processes a
bounded set of views at a prescribed input resolution. Applying it to
high-resolution or large-view reconstruction therefore typically requires
restricting the number of views or resizing the images, which may discard
fine-scale visual evidence. Such information loss is especially detrimental
to thin structures, sharp depth discontinuities, and small geometric details.
By operating directly on native-resolution patches, SPGE avoids the accuracy
degradation caused specifically by mandatory downsampling. Thus, although its
runtime increases with the number of input images, SPGE provides a favorable
trade-off between computational cost, input flexibility, and reconstruction
quality.

\subsection{Post-processing Compatibility on DTU}
\label{subsec:supp_postprocessing}

\begin{figure*}[t]
    \centering
    \includegraphics[width=\textwidth]
    {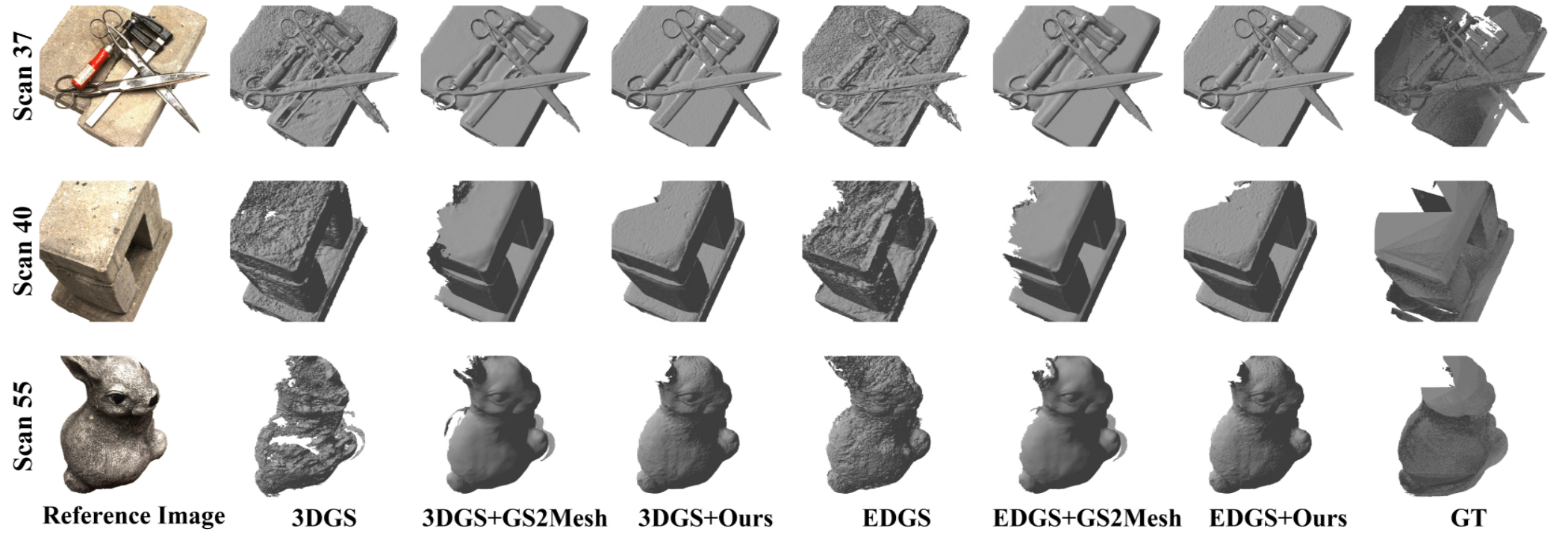}
    \caption{Compatibility of \method{} as a post-processing module on DTU.
    Each row presents results obtained from either a standard 3DGS or an EDGS
    representation. The input Gaussian representations remain fixed during
    mesh reconstruction.}
    \label{fig:supp_postprocessing_dtu}
\end{figure*}

We evaluate \method{} as a general post-processing module for Gaussian
representations produced by different optimization pipelines. Specifically,
we apply the same trained model to standard 3DGS and
EDGS~\cite{kotovenko2026edgs} scenes on DTU~\cite{jensen2014dtu}
without updating the input Gaussian primitives or performing scene-specific
fine-tuning. As shown in
Figure~\ref{fig:supp_postprocessing_dtu}, the proposed method consistently
removes floating geometry, fills incomplete surfaces, and improves geometric
continuity while preserving the appearance and fine structures encoded by
the original representation. 

\subsection{Qualitative Comparison with Gaussian-to-Mesh Methods}
\label{subsec:supp_method_comparison}

\begin{figure*}[t]
    \centering
    \includegraphics[width=\textwidth]
    {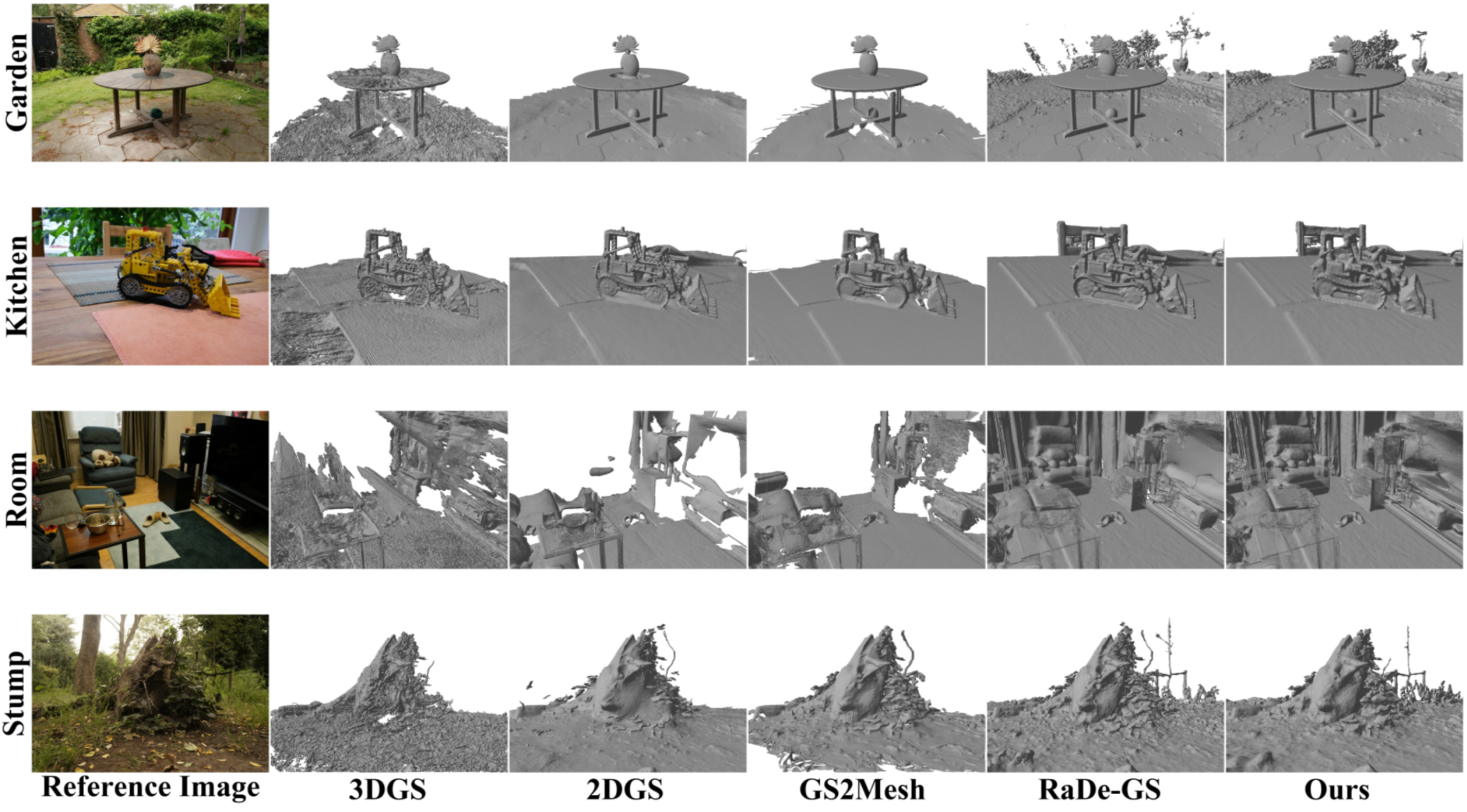}
    \caption{Qualitative comparisons on the Mip-NeRF 360 dataset (Garden,
    Kitchen, Room, and Stump). We compare \method{} with 3DGS, 2DGS,
    GS2Mesh, and RaDe-GS. Without scene-specific mesh optimization, \method{}
    generalizes to unbounded real-world scenes and reconstructs cleaner, more
    spatially coherent surfaces while preserving foreground structures and
    complex backgrounds.}
    \label{fig:supp_qualitative_mipnerf360}
\end{figure*}

Figures~\ref{fig:supp_qualitative_mipnerf360}
and~\ref{fig:supp_qualitative_dtu} present additional comparisons with
representative Gaussian-to-mesh and Gaussian-based surface reconstruction
methods on Mip-NeRF 360~\cite{barron2022mipnerf360} and
DTU~\cite{jensen2014dtu}, respectively. On DTU,
where ground-truth geometry is available, \method{} produces more complete
surfaces and preserves thin structures and depth discontinuities more
accurately. The Mip-NeRF 360 results further demonstrate that the trained
model generalizes to unbounded real-world scenes without scene-specific mesh
optimization.

Compared with GS2Mesh~\cite{wolf2024gs2mesh}, the proposed feed-forward reconstruction avoids
iterative stereo-depth fusion and produces more spatially consistent
geometry. Compared with surface-oriented Gaussian methods such as
SuGaR~\cite{guedon2024sugar}, 2DGS~\cite{huang2024twodgs},
RaDe-GS~\cite{zhang2024radegs}, QGS~\cite{zhang2025qgs},
PGSR~\cite{chen2024pgsr}, and GGGS~\cite{zhang2026geometry}, \method{} provides competitive geometric
quality while operating directly as a post-processing module for an existing
Gaussian representation.

\begin{figure}[t]
    \centering
    \includegraphics[width=\linewidth]
    {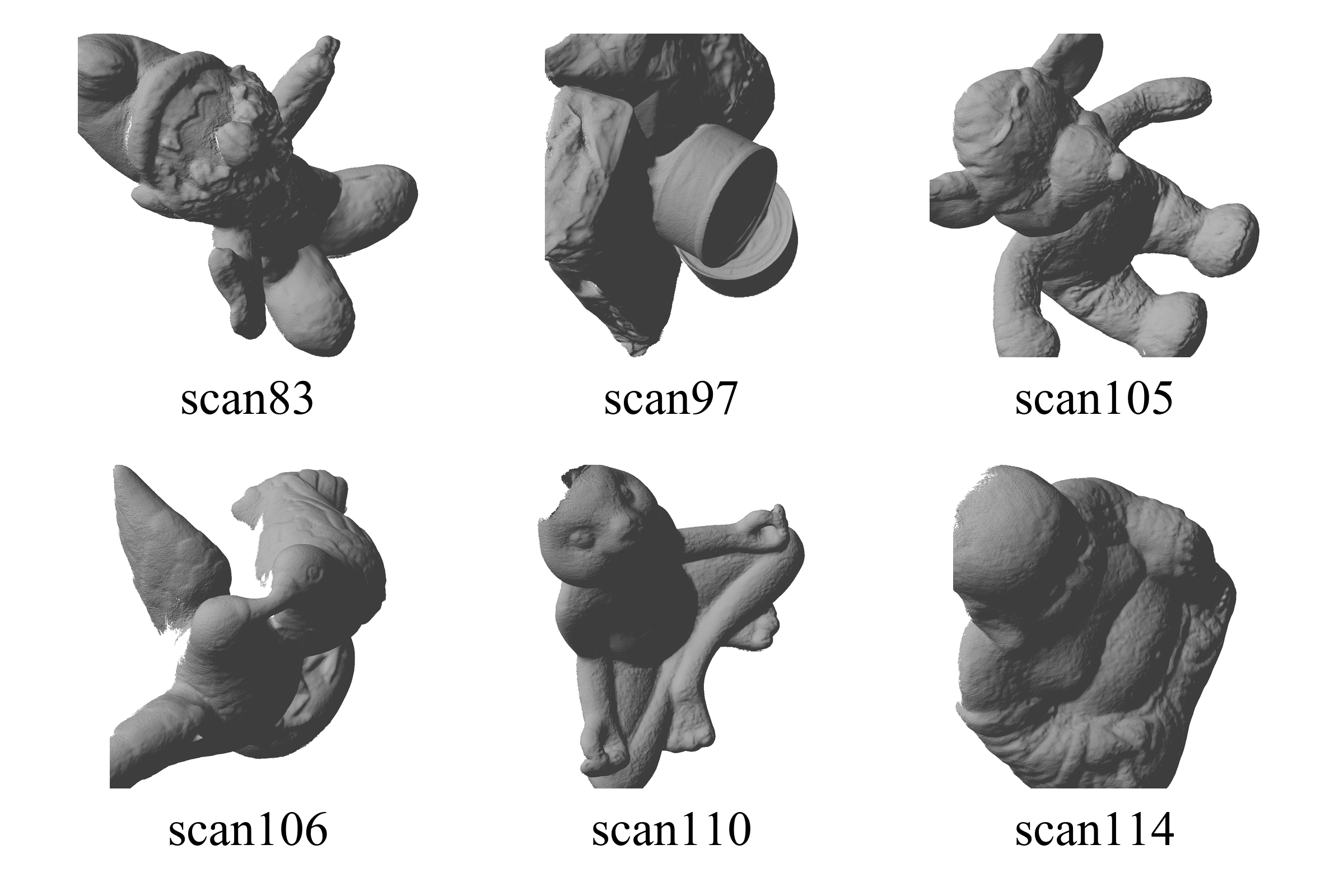}
    \caption{Qualitative Results of Our Method on the DTU Dataset}
    \label{fig:scan83_97_105_106_110_114_overall}
\end{figure}

Figure~\ref{fig:scan83_97_105_106_110_114_overall} presents additional
qualitative results of \method{} on six DTU scenes.

\subsection{Analysis of Intermediate Depth and Geometry Predictions}
\label{subsec:scan83_97_105_106_110_114_overall}

\begin{figure*}[p]
    \centering

    \includegraphics[width=0.98\textwidth]
    {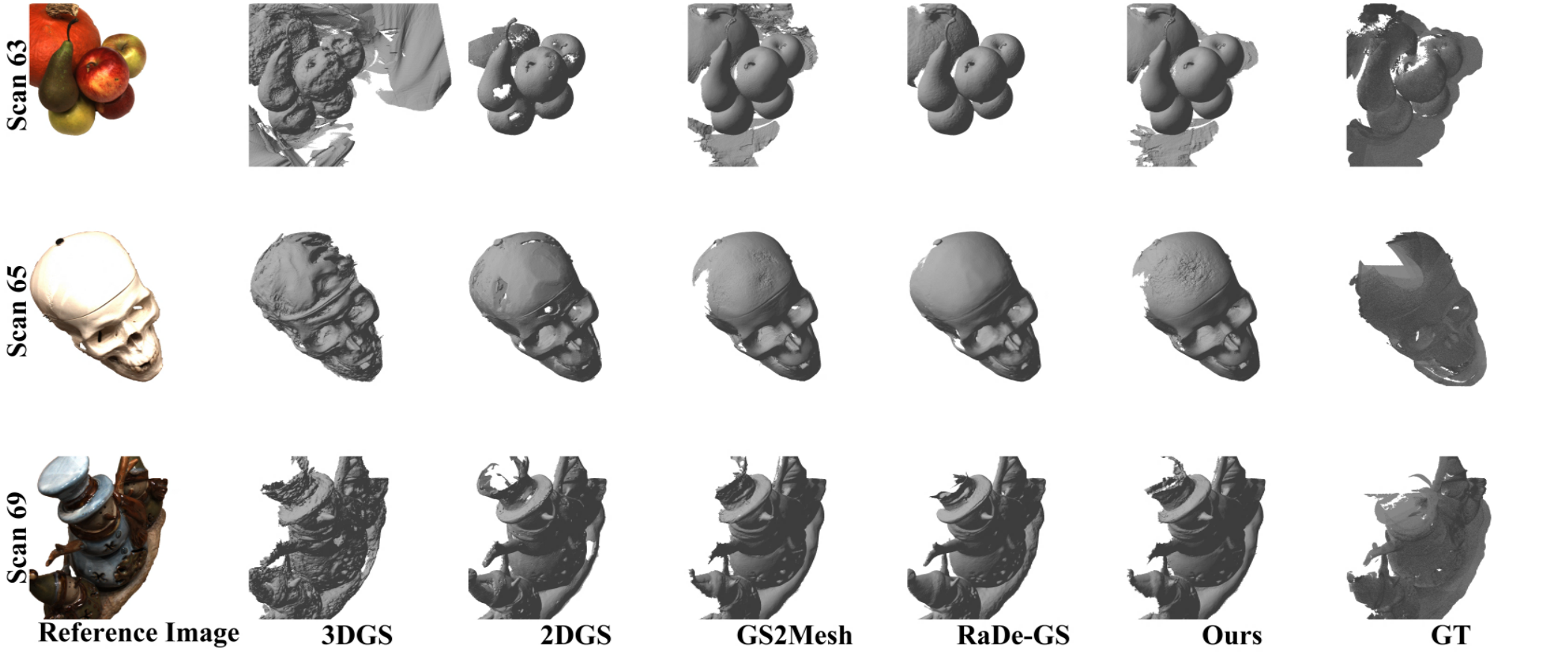}
    \caption{Qualitative comparisons on DTU scans 63, 65, and 69. We compare
    \method{} with 3DGS, 2DGS, GS2Mesh, and RaDe-GS; ground-truth geometry is
    shown in the last column. \method{} recovers more complete surfaces,
    suppresses floating artifacts, and better preserves thin structures and
    sharp depth discontinuities, producing geometry that more closely matches
    the ground truth.}
    \label{fig:supp_qualitative_dtu}

    \vspace{0.4em}

    \includegraphics[width=0.98\textwidth]
    {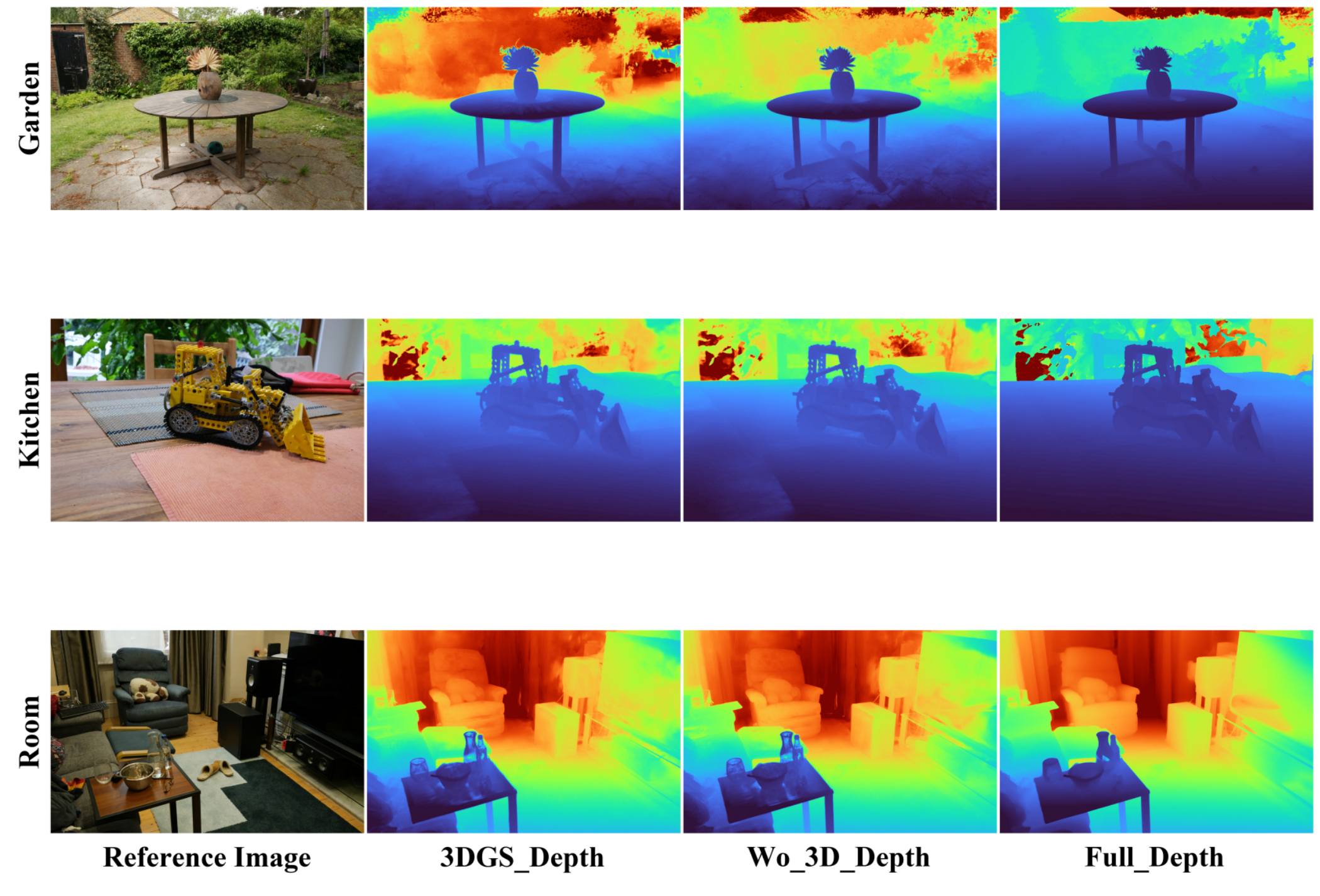}
    \caption{Analysis of intermediate depth predictions on Mip-NeRF 360.
From left to right, we show the reference image, the metric depth rendered
from the input 3DGS representation (\textit{3DGS\_Depth}), the prediction
obtained without the 3D Gaussian stream (\textit{Wo\_3D\_Depth}), and the
refined depth produced by the full model (\textit{Full\_Depth}). The visual
branch recovers image-aligned structures but may lack stable metric and
cross-view constraints. Incorporating the Gaussian stream, Visual--Gaussian
spatial reasoning, and scale-aligned hybrid refinement suppresses floating
artifacts, improves foreground-background separation, and produces more
spatially coherent geometry.}
    \label{fig:supp_intermediate_depth}
\end{figure*}

To better understand how the proposed modules progressively transform a
rendering-oriented Gaussian representation into coherent surface geometry, we
visualize several intermediate depth predictions in
Figure~\ref{fig:supp_intermediate_depth}. These results illustrate the
complementary roles of the native-resolution visual stream, the 3D Gaussian
stream, Visual--Gaussian spatial reasoning, and the final scale-aligned depth
refinement.

The initial depth $D_i^0$, denoted as \textit{3DGS\_Depth} in the figure, is
rendered directly from the input 3DGS representation. It provides a metric
geometric reference because it shares the coordinate system of the input
Gaussian scene. However, since the original Gaussian primitives are optimized
primarily for novel-view rendering rather than explicit surface
reconstruction, their depth maps often contain floating structures, noisy
background geometry, incomplete surfaces, and inaccurate depth boundaries.
These artifacts are particularly visible around thin objects, occlusion
boundaries, and sparsely observed regions.

The \textit{Wo\_3D\_Depth} results are produced after removing the 3D Gaussian
stream while retaining the native-resolution visual processing of SPGE. The
visual branch exploits high-resolution RGB evidence to recover object
silhouettes, local structures, and image-aligned depth discontinuities.
However, without explicit features extracted from the scene-level Gaussian
representation, its prediction lacks a sufficiently strong metric and
cross-view geometric constraint. Consequently, the estimated depth may appear
locally plausible while remaining less stable in textureless regions, distant
backgrounds, and areas with ambiguous visual evidence.

In the full model, the PTV3-based Gaussian stream first extracts geometric
features from the input Gaussian primitives. GSRT then projects and associates
these features with their corresponding visual patches, enabling each image
token to reason jointly about native-resolution appearance and explicit 3D
geometry. The visual decoder predicts an image-conditioned depth
$D_i^{\mathrm{2D}}$, which preserves fine structures and sharp boundaries,
while the Gaussian decoder predicts surface-aware Gaussian primitives from
which the metric depth $D_i^{\mathrm{3D}}$ is rendered. The former provides
high-frequency local evidence, whereas the latter provides scene-level metric
structure and stronger cross-view consistency.

Finally, SHDR aligns $D_i^{\mathrm{2D}}$ with the metric scale of
$D_i^{\mathrm{3D}}$ and combines the complementary predictions to obtain the
refined depth $D_i^*$, shown as \textit{Full\_Depth}. In the Garden example,
the full model produces a clearer separation between the table, foreground
objects, and distant background. In the Kitchen example, it preserves the
loader silhouette while suppressing unstable background geometry. In the Room
example, it recovers more coherent furniture boundaries and smoother planar
structures. These comparisons demonstrate that visual evidence alone is
insufficient for globally consistent reconstruction: explicitly incorporating
the Gaussian stream and scale-aligned hybrid refinement is essential for
simultaneously preserving local image details and maintaining coherent metric
geometry across views.

\bibliography{aaai2027}


\end{document}